\documentclass[%
  jmp,
 amsmath,amssymb,
preprint,11pt,%
]{revtex4-1}

\usepackage{graphicx}
\usepackage{dcolumn}
\usepackage{bm}

\usepackage[utf8]{inputenc}
\usepackage[T1]{fontenc}
\usepackage{enumitem}
\usepackage{mathptmx}

\newcommand{\e}{\text{e}}
\newcommand{\de}{\text{d}}
\newcommand{\Tr}{\text{Tr}}

\begin{document}

\title{Noise Effects on the Wilczek-Zee Geometric Phase}

\author{Pedro Aguilar}
\email{pedro.aguilar@nucleares.unam.mx}
\author{Chryssomalis Chryssomalakos}
\email{chryss@nucleares.unam.mx}
\author{Edgar Guzm\'an-Gonz\'alez}
\email{edgar.guzman@correo.nucleares.unam.mx}
 \altaffiliation[Current address:]{
 London Mathematical Laboratory, 18 Margravine Gardens, London W6 8RH, United Kingdom.
 }
\affiliation{Instituto de Ciencias Nucleares\\ Universidad Nacional Aut\'onoma de M\'exico \\
	PO Box 70-543, 04510, CDMX, M\'exico}

\date{\today}

\begin{abstract}
 \noindent 
Non-abelian geometric phases have been proposed as an essential ingredient in logical gate implementation --- their geometric nature guarantees their invariance under reparametrizations of the associated cyclic path in parameter space. However, they are still dependent on deformations of that path, due to, \emph{e.g.}, noise. The first question that we tackle in this work is how to quantify in a meaningful way this effect of noise, focusing, for concreteness, on the nuclear quadrupole resonance hamiltonian --- other systems of this nature can clearly be treated analogously. We consider a precessing magnetic field that drives adiabatically a degenerate doublet, and is subjected to noise, the effects of which on the Wilczek-Zee holonomy are computed analytically. A critical review of previous related works reveals a series of assumptions, like sudden jumps in the field, or the presence of white noise, that might violate adiabaticity. We propose a state-independent measure of the effect, and then consider sinusoidal noise in the field, of random amplitude and phase. We find that all integer noise frequencies $m\neq 2$ behave similarly, in a manner reminiscent of the abelian case, but that noise of frequency $m=2$ has a very different, and, at the same time, very pronounced effect, that might well affect robustness estimations.
\end{abstract}

\maketitle
\section{Introduction}
\noindent Recent interest in applications of geometric phases to quantum computing~\cite{Zanardi199994, pachos2001quantum,292c4cfc0e4847d9aeb6240fd0d707b8}, based on the seminal work of Berry~\cite{Berry08031984}, Wilczek and Zee~\cite{PhysRevLett.52.2111}, and Aharonov and Anandan~\cite{PhysRevLett.58.1593,Anandan1988171}, has prompted 
a flurry of related activity in the last decade or so.

A natural division of the above work is into the abelian and non-abelian case. In the first, which is the only one where the term ``phase'' applies literally, the wavefunction of a non-degenerate hamiltonian eigenstate acquires a phase factor, additional to the expected dynamical one, every time it traces adiabatically a loop in the projective Hilbert space $P(\mathcal{H})$ of the system. In the second, the eigenstate is assumed degenerate, and the above phase factor gets promoted to a unitary matrix operating in the degenerate subspace. In both cases, the ``phase'' only depends on the loop traced in $P(\mathcal{H})$, and is insensitive to its parametrization, hence the term ``geometric''.  
Generalizations exist that drop both the eigenstate and the adiabatic requierement~\cite{PhysRevLett.60.2339,0305-4470-23-11-027,PhysRevA.76.044303,Xu2012}, but we work with the original formulation as the distinction between the two is not essential for our purposes. 

A second possible division, transversal to the above, is into experimental, numerical, and theoretical approaches.
 On the experimental front, a multitude of devices, including Josephson junctions and other superconducting circuits~\cite{mottonen2008experimental,brosco2008non,faoro2003non,solinas2010ground,pirkkalainen2010non}, NMR systems~\cite{Feng2013,Li2017}, electronic harmonic oscillators~\cite{pechal2012geometric}, trapped ions~\cite{duan2001geometric}, multi-level atoms~\cite{unanyan1999laser,AbdumalikovJr2013}, solid state defects~\cite{ArroyoCamejo2014,Zu2014} and electron spins in quantum dots~\cite{golovach2010holonomic} or interacting with light~\cite{Sekiguchi2017}, among others, have already been considered as candidates for quantum logical gates. On the other hand, 
much theoretical effort has been expended evaluating the potential performance of such devices, with particular attention to their robustness. Potential degrading factors for quantum computing are generally divided in enviromental ones, that lead, typically, to decoherence~\cite{PhysRevLett.90.160402,PhysRevLett.90.190402,1751-8121-44-9-095304,FuentesGuridi2005,Thunstrom2005,Sarandy2006,Moller2008}, and parametric ones, related to the finite accuracy of parameter manipulation~\cite{She.Bro.Wha:03,deChiara:2008,PhysRevLett.91.090404,solinas2004robustness,solinas2012stability}. The seminal work of Shor~\cite{Sho:94} and Grover~\cite{Gro:96}, whose algorithms catapulted quantum computing to the forefront of research, has provided a concrete testing ground for the effect of gate degradation on algorithmic complexity --- detailed analyses  
can be found in~\cite{Lon.Li.Zha.Tu:00,Guo2001}. Lying somewhere in between and bridging the above two fronts, numerical simulations have also been carried out, modelling both decoherence~\cite{Niwa2002,Salas2007} and  parametric noise~\cite{Niwa2002,She.Bro.Wha:03,filipp:2008}, followed by experiments designed to test their results~\cite{PhysRevLett.102.030404,PhysRevA.87.060303}.  

Our own motivation to contribute to this line of research stemmed from the analysis of parametric noise effects in the archetypical system of a spin-1/2 coupled to a precessing magnetic field contained in~\cite{PhysRevLett.91.090404}. In that work, a stochastic component, simulating noise, is added to the field, and the resulting average geometric  phase is computed, along with its standard deviation, to first order in the noise amplitude. 
The answer for the average phase is, predictably, null, as the noise is assumed to have  zero mean, and the analysis is linear in the noise components. Two natural questions that we posed were to calculate, on the one hand, the first non-vanishing correction to the phase, expected to be quadratic in the noise, and, in a more conceptual direction, consider a \emph{quantum} vector noise, and compare to the classical case --- our conclusions are contained in~\cite{Agu.Chr.Guz:15a}. The next natural step, it seems to us, is to explore what happens in the non-abelian case. Here, too, one may consider classical stochastic noise, and inquire about the statistics of the resulting holonomies, to the first non-vanishing order in the noise, or, more ambitiously, one may envisage quantum fluctuations of the ``parameters'', which would now be themselves quantum dynamical variables, and compute corrections to the geometric phase due to quantum fluctuations in the ``parameters''. The present work  deals with the first of these problems, in the particular example of nuclear quadrupole resonance, that has been studied both theoretically~\cite{PhysRevA.38.1} and experimentally~\cite{Tycko1987} (note though that in this last reference, the treatment is essentially abelian). Although this system is not usually associated to quantum computing, it serves well as a prototype for this kind of analysis --- we expect our methods and conclusions to be easily adaptable to more realistic scenarios.

Some related work has appeared before,  in~\cite{solinas2004robustness,solinas2012stability}. In the former, a numerical approach is adopted. The control parameters $\Omega_i$, $i=1,2,3$, of a hamiltonian are perturbed by piecewise constant noise, that changes after a fixed time interval $\tau$, so that the (3D) vector parameter $\vec{\Omega}(t)$ takes the value $\vec{\Omega}_0+\delta\vec{\Omega}_k$ for $k \tau \leq t < (k+1)\tau$, $k \in \mathbb{Z}$. The associated holonomy only depends on the area enclosed by the curve the direction of $\vec{\Omega}$ traces on the unit sphere. The presence of noise modifies this curve, and the modulus of the  inner product of the resulting state with the unperturbed one (known as \emph{fidelity}) is used as an indicator of the effect of noise. The result is averaged over a sample of 18 states, and over five ``runs'' of the noise, each run producing a particular value for $\delta\vec{\Omega}_k$, according to a fixed gaussian distribution.

There are two aspects of the above approach that we find worrisome: first, the sudden jumps of the noise at integer multiples of $\tau$ might well violate the adiabaticity condition. There is, in fact, ample evidence that this indeed happens, in Figure 7 of that reference, where the non-logical states become heavily populated, a clear sign of non-adiabatic transitions taking place. Second, for the averaging over states a particular (somewhat symmetric, but otherwise arbitrary) grid on the unit sphere was chosen, without any quantification of the dependence of the results on this choice. These considerations added two items to our to-do list: (i) work with smooth sinusoidals as noise, limiting their frequency so that adiabaticity is preserved, and (ii) find a way to quantify the effect of noise that does not depend on any choice of states.

Ref.~\cite{solinas2012stability}, on the other hand, takes an analytical approach, which would make it, in principle, more directly comparable with our intended analysis. After working out the holonomy in the presence of noise, the authors of~\cite{solinas2012stability} turn their attention to its statistics, choosing, in their Eq.~(50), a flat-spectrum white noise. It is not clear to us, however, whether this choice respects adiabaticity, as the high frequencies (with respect to the scale dictated by the spectrum of the hamiltonian) are not adequately suppressed: no physical system can lock adiabatically onto white noise.  On the positive side, instead of dubiously averaging over arbitrarily selected states, the authors of~\cite{solinas2012stability} choose to focus on the holonomy, and quantify noise effects in a state-independent way. Accordingly, our to-do list gets two more entries: (iii) define the noise \emph{statistics} in a way consistent with adiabaticity, and (iv) focus on the holonomy itself, rather than on its output acting on particular states.

Given this state of affairs, we revisit the subject here, benefiting, on the one hand, from all that \emph{is} done correctly in the above two references, but trying, at the same time, to improve on the above mentioned aspects of the problem. A final point that we felt deserved some attention was that the hamiltonian considered in both of the above works gives rise to a rather mildly non-abelian holonomy: as the authors of the second reference point out (right after their Eq.~(35)), their Wilczek-Zee connection matrix $\mathcal{A}(t)$ is such that $[\mathcal{A}(t_1),\mathcal{A}(t_2)]=0$, for all $t_1$, $t_2$, so that the time evolution is given by simple exponentiation. This is reflected in the fact that the resulting holonomy is proportional to the area on the unit sphere of $\hat{\Omega}$ described above, a hallmark feature of the abelian case, that has no \emph{a priori} reason to persist in a genuinely non-abelian case (in fact, it does not, as we show in what follows). To avoid such oversimplifications, we consider, as mentioned above, nuclear quadrupole resonance, in which a spin-3/2 couples to a magnetic field in such a way that two degenerate doublets form. The Wilczek-Zee connection in one of them is diagonal, but turns out truly non-abelian in the other --- this is the space our qubit lives in, and our single-qubit gate operates on. 
Our aim in this paper is to properly formulate the problem of noise effects on holonomies, and present reliable results of the resulting statistics, with an eye towards eventual applications in quantum logical gates. For this reason, we do not consider two-qubit gates, which, despite being necessary for universal quantum computing, present little novelty from our present point of view. Relevant general references  that we have found helpful include~\cite{messiah1962,peres:1995,bohm:2003,dariusz:2004,
jacobs:2010,Nak:90,Bengtsson2008}.

We conclude this introduction with an overview of the rest of the paper.  In section \ref{geometricPhases}, we offer a short presentation of adiabatic non-abelian geometric phases and of their most relevant properties for the following sections, as well as the standard treatment of nuclear quadrupole resonance. In section \ref{NQRNoisy} we present our treatment of the nuclear quadrupole resonance when the driving magnetic field is fluctuating randomly, to the first non-trivial order in the noise amplitude. We employ a measure of these effects that is defined in the space of unitary matrices themselves, unlike the commonly employed Bures or Fubini-Study metrics. We study the effects of individual Fourier modes of fluctuations and find the corresponding probability distribution of the holonomy. Finally, in section \ref{conclusions} we present conclusions, final remarks and ideas for future research.
\section{Wilczek-Zee geometric phases}
\label{geometricPhases}
\noindent We restrict ourselves to the adiabatic cyclic formulations of the non-abelian geometric phase, namely, we present here the relevant aspects of the  Wilczek-Zee \cite{PhysRevLett.52.2111} geometric phase.

Consider a physical system described by a hamiltonian $H(\xi)$ that depends on external parameters $\xi^i$, the spectrum of which presents degeneracy. Let $\mathcal{H}_n^D[\xi]$ be the $D$-dimensional degenerate subspace corresponding to the $n$-th energy level of the hamiltonian $H(\xi)$, with basis elements  $|a,\xi\rangle$ for $a=1,\ldots,D$. Suppose that the $\xi$'s vary in time adiabatically, tracing a curve $C \colon \tau \rightarrow \xi^i(\tau)$ in the parameter space $M$ --- through these parameters, the hamiltonian acquires a time dependence. When the initial state is degenerate, and this degeneracy is preserved under adiabatic evolution, nothing prevents the state from mixing with other states with the same energy --- this mixing is described by a connection matrix $A^{(D)}$, with 1-form entries given by
\begin{equation}\label{WZEntries}
A^{(D)}_{ab} = i\langle a,\xi|\de|b,\xi\rangle
\, ,
\end{equation}
where $a$, $b$ range over degenerate eigenstates. $A^{(D)}$ is commonly referred to as the Wilczek-Zee connection because it transforms as a non-abelian $\text{U}(n)$ connection under local unitary transformations of the degenerate subspace. It is also evident from (\ref{WZEntries}) that $A^{(D)}$ is parameterization independent. The initial state, after a complete cycle $C$ of the parameters, will acquire a dynamical phase $\gamma_d$ and will get mixed with elements belonging to its degenerate subspace via the unitary matrix
\begin{equation}
U_C^{(D)} = \mathcal{P}\e^{i\oint_C A^{(D)}}
\, ,
\end{equation}
\emph{i.e.}, the path ordered integral of the Wilczek-Zee connection along the curve $C$. Note that if in some basis of the degenerate subspace the Wilczek-Zee connection is diagonal, then, in that basis, no mixing in the subspace occurs and each degenerate state just accumulates its corresponding Berry's phase.

A physical system exhibiting non-trivial Wilczek-Zee geometric phases is nuclear quadrupole resonance \cite{PhysRevA.38.1}. Consider the hamiltonian $H = \mu(\mathbf{B}\cdot\mathbf{J})^2$ describing the coupling of a spin-3/2 particle with an external magnetic field $\mathbf{B}$. The eigenstates of this hamiltonian are spin-3/2 states with spin projection $m$ along the $\hat{\mathbf{B}}$ direction, which we denote by $|\hat{\mathbf{B}},m\rangle$. The pair of states with projections $+m$ and $-m$ are degenerate, therefore there exist two 2-dimensional degenerate subspaces, corresponding to $m=\pm 1/2$ and $m=\pm 3/2$. When the magnetic field traces adiabatically some loop, the evolution of degenerate eigenstates will correspond to some non-trivial mixing within their subspaces, given by the path ordered exponential of the Wilczek-Zee connection. The space of parameters in this case is the 2-sphere on which $\hat{\mathbf{B}}$ lives. The connection in the $m=\pm 3/2$ subspace is abelian, that is, there is no mixing between states because $\mathbf{J}$ can only produce transitions with $\Delta m = 0,\pm 1$. Only in the $m=\pm 1/2$ subspace a non-trivial mixing occurs.

For the computation of the Wilczek-Zee connection we will need the instantaneous eigenstates of $H$, parameterized by coordinates on the 2-sphere, the local character of the latter giving rise to subtleties that need special care~\cite{PhysRevA.38.1}.  We obtain the eigenstates of $H$ by rotation of the eigenstates of $J_z$, written as vectors $(1,0,0,0)$, and so on, in column vector form, along the geodesic connecting the north pole to the point with standard spherical  coordinates $(\Theta,\Phi)$ defining the $\hat{\mathbf{B}}$ direction. The corresponding rotation axis and angle are $(-\sin\Phi,\cos\Phi,0)$ and $\Theta$, respectively, so that the rotation in state space is effected by $\exp(-i\Theta(-\sin\Phi,\cos\Phi,0)\cdot\mathbf{J})$ and the resulting $H$-eigenstates are
\begin{equation}
\label{eigenpm}
|\hat{\mathbf{B}},+1/2\rangle = 
\begin{pmatrix}
 -\frac{\sqrt{3}}{4}\e^{-i\Phi}\sin^2\Theta\csc(\Theta/2)\\
 \frac{1}{4}\left[\cos(\Theta/2)+3\cos(3\Theta/2)\right]\\
 -\frac{1}{4}\e^{i\Phi}\left[\sin(\Theta/2)-3\sin(3\Theta/2)\right]\\
 \frac{\sqrt{3}}{2}\e^{2i\Phi}\sin\Theta\sin(\Theta/2)
\end{pmatrix},
\quad
|\hat{\mathbf{B}},-1/2\rangle = 
\begin{pmatrix}
 \frac{\sqrt{3}}{2}\e^{-2i\Phi}\sin\Theta\sin(\Theta/2)\\
 \frac{1}{4}\e^{-i\Phi}\left[\sin(\Theta/2)-3\sin(3\Theta/2)\right]\\
 \frac{1}{4}\left[\cos(\Theta/2)+3\cos(3\Theta/2)\right]\\
 \frac{\sqrt{3}}{4}\e^{i\Phi}\sin^2\Theta\csc(\Theta/2)
\end{pmatrix}.
\end{equation}
Using (\ref{WZEntries}), the Wilczek-Zee connection in this gauge is
\begin{equation}\label{WZConnectionNorthGauge}
A_N = \left(-\sin\Phi\sigma_1+\cos\Phi\sigma_2\right)\de\Theta + \left(-\sin\Theta\cos\Phi\sigma_1-\sin\Theta\sin\Phi\sigma_2+\frac{1}{2}(\cos\Theta-1)\sigma_3\right)\de\Phi.
\end{equation}
The non-abelian character of the Wilczek-Zee holonomy shows itself in that integrating this connection along a closed path on the sphere gives rise, in general, to a nontrivial unitary matrix.

We note that for a constant-$\Theta$ loop close to the the north pole, $A_N\approx 0$, while near the south pole we have $A_N\approx -\sigma_3\de\Phi$, so that, even for an infinitesimal loop, the resulting holonomy is a $2\pi$ rotation around $\hat{z}$. This connection is then defined on the sphere excluding a region around the south pole (the north patch). A good connection $A_S$ for the south patch is obtained after a gauge transformation $A_S = \rho_1^\dagger A_N\rho_1 + i\rho_1^\dagger\de\rho_1$ with $\rho_1 = \e^{-i\Phi \sigma_3}$, resulting in
\begin{equation}
A_S = \left(\sin\Phi\sigma_1+\cos\Phi\sigma_2\right)\de\Theta + \left(-\sin\Theta\cos\Phi\sigma_1+\sin\Theta\sin\Phi\sigma_2+\frac{1}{2}(\cos\Theta+1)\sigma_3\right)\de\Phi.
\end{equation}
For calculational purposes, the form of these connections is not convenient. Restricting to the north patch, through the gauge transformation $A_E=\rho_2^\dagger A_N\rho_2 + i\rho_2^\dagger\de\rho_2$ with a locally defined $\rho_2 = \e^{-i\Phi\sigma_3/2}$, we obtain the connection
\begin{equation}\label{WZConnectionEquator}
A_E = \sigma_2\de\Theta + \left(-\sin\Theta\sigma_1+\frac{1}{2}\cos\Theta\sigma_3\right)\de\Phi,
\end{equation}
which is constant along segments with constant $\Phi$ or $\Theta$ and therefore generates rotations along fixed axes that can be immediately integrated. The path ordered exponentials of $A_N$ and $A_E$, which we call $U_N$ and $U_E$, respectively, are related by $U_N=\rho_2 U_E = \e^{-i\Phi\sigma_3/2}\mathcal{P}\e^{i\int A_E}$. Therefore, for the closed loop $C_0$ consisting of $\Theta$ fixed at some value $\Theta_0$ and $0\leq\Phi<2\pi$, the holonomy is
\begin{equation}
\label{U0def}
U_\text{N}(C_0) \equiv  U_0 = \e^{-i\pi}\exp\left(2i\pi \left(-\sin\Theta_0\sigma_1+\frac{1}{2}\cos\Theta_0\sigma_3\right)\right).
\end{equation}

\vspace{1cm}
\noindent In the following sections, we will apply these results for nuclear quadrupole resonance when the driving parameters have a fluctuating component.
\section{Nuclear quadrupole resonance driven by a fluctuating magnetic field}
\label{NQRNoisy}
\noindent In the standard example of Berry's phase, involving a spin-1/2 coupled to a precessing magnetic field, the effect of stochastic fluctuations of the field has been analyzed in \cite{PhysRevLett.91.090404}. The approach taken there was to include a fluctuating component in the magnetic field, and compute Berry's phase as minus one half the solid angle subtended by the curve traced out by the magnetic field. We propose a more economical treatment, based on the geometrical nature of the Wilczek-Zee phase: since all that matters in the calculation is the curve itself, not its particular parametrization, we opt to skip the particulars of how it was produced by the fluctuating field, and focus instead on its form, given as a function $\Theta(\Phi)$. In this way, the considerable complexity introduced in the translation of fluctuating field components into curve deformations is bypassed, and the calculation simplifies considerably. 

\begin{figure}[h]
\hfill
\includegraphics[width=.85\linewidth]{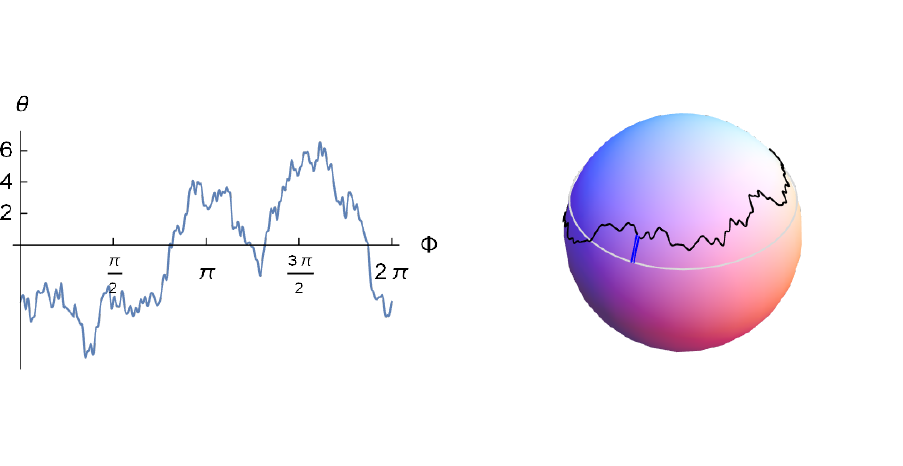}
\hfill
\caption{Left: a particular realization of $\theta(\Phi)=\theta_0/\sqrt{2\pi}+\sum_{k=1}^{100} (\theta_k^* e^{-ik\Phi} +\theta_k e^{ik \Phi})/\sqrt{2\pi}$, where $\theta_0$,  $\Re(\theta_k)$, $\Im(\theta_k)$ are taken to be normally distributed, with zero average and standard deviation $\sigma_0=\sqrt{2\pi}$, $\sigma_k=\sqrt{2\pi}/k$, $k=1,\ldots,100$. Note that $\theta(\Phi)$ is periodic  but does not necessarily vanish at its end points. Right: the corresponding curve traced by the magnetic field direction on the unit sphere --- the relatively large value of $\epsilon=0.05$ was used for visual clarity. The fluctuating curve $\Theta(\Phi)=\Theta_0+\epsilon \theta(\Phi)$, with $\Theta_0=\pi/3$,  is drawn in black, the unperturbed curve $\Theta(\Phi)=\Theta_0$ in gray, and the initial and final ``closing'' segments (shown slightly separated for clarity) in blue.}
\label{fig:fluctuations}
\end{figure}

The noisy curves we consider are infinitesimal deformations of the constant-$\Theta$ circles describing a precessing field, \emph{i.e.}, we take $\Theta(\Phi)=\Theta_0+\epsilon \theta(\Phi)$, where $\theta(\Phi)$ is a periodic stochastic process, $\theta(0)=\theta(2\pi)$, with values of order 1. We do not assume $\theta(0)=0$, meaning that while the initial  degenerate subspace coincides with the final one, in each realization of $\theta(\Phi)$, they differ among themselves from one realization to another. This renders the naive averaging of the corresponding holonomies meaningless, as each one of them operates in a different Hilbert space. We remedy the situation by fixing the base point $(\Theta_0,0)$ on the unperturbed curve, and considering infinitesimal segments that connect it to the initial (and final) point of the perturbed curve. Analytically, we define
\begin{equation}
\label{ThetaPhidef}
(\Theta(t),\Phi(t)) = 
\begin{cases}
	(\Theta_0+\epsilon\theta(0)(t+1),0) & \mbox{if } -1\leq t<0\\
	(\Theta_0+\epsilon\theta(t),t) & \mbox{if } 0\leq t<2\pi \\
	(\Theta_0 + \epsilon\theta(0)-\epsilon\theta(0)(t-2\pi),2\pi) & \mbox{if } 2\pi\leq t<2\pi+1 \\
\end{cases}
\end{equation}
where the first and third segments are of order $\epsilon$, their purpose being to make all realizations depart from and return to the same base point. 
The curve describes  a circle of fixed polar angle $\Theta_0$ subject to random fluctuations given by $\theta(\Phi)$ at each value of the azimuthal angle $\Phi$. The parameter $\epsilon$ characterizes the size of the fluctuations and is assumed to be much smaller than one (the radius of the sphere of magnetic field directions). To arrive at concrete numerical results, one needs to specify the particular statistical properties of $\theta(\Phi)$, which we do later on. An example of such statistics is used in Figure \ref{fig:fluctuations} to produce a particular realization of $\theta(\Phi)$, and the curve it corresponds to on the unit sphere.

 Note that our assumption above of a periodic noise $\theta(\Phi)$, motivated by the technical convenience of a discrete spectrum,  and which, admittedly, might seem somewhat artificial, could be easily relaxed. In that case, the infinitesimal segments connecting $\Theta(0)$, $\Theta(2\pi)$ to the unperturbed curve would be, in general,  unequal --- it is easily seen though that this would not affect our results to leading order: in the non-periodic case the curve traced out by the magnetic field would not close nicely onto itself, but it could be modified smoothly, in a $\Phi$- interval of order $\epsilon$, $(2\pi-\epsilon,2\pi)$, so that it does --- such modification, the leading effect of which would vanish on the average, would only affect higher-order corrections. Having said that, we also point out that \emph{some} (but certainly not all) sources of noise are, by nature, periodic: for example, the effects of any background static stray magnetic field present in the lab, that simply adds to the rotating one, or those of mechanical imperfections in the rotating apparatus, resulting, \emph{e.g.}, in a slightly elliptical path for the magnetic field, may be treated as periodic (albeit not necessarily stochastic) noise.

We continue our calculation using $A_E$ in (\ref{WZConnectionEquator}) as the Wilczek-Zee connection. Using Eq.~(\ref{ThetaPhidef}) we compute $\de \Theta$, $\de \Phi$ along the three segments of the curve  --- expanding everything up to $\mathcal{O}(\epsilon^2)$ we obtain for the connection
\begin{equation}\label{WZConnectionWholeCurve}
A_E(t) = 
\begin{cases}
\epsilon\theta(0)\sigma_2 \,  \de t & \mbox{if } -1\leq t<0\\
(A_0  + \epsilon A_1  + \epsilon^2 A_2) \, \de t + \mathcal{O}(\epsilon^3) & \mbox{if } 0\leq t<2\pi \\
-\epsilon\theta(0)\sigma_2 \, \de t & \mbox{if } 2\pi\leq t<2\pi+1 \\
\end{cases}
\end{equation}
with
\begin{equation}\label{WZConnectionMiddleCurve}
\begin{aligned}
A_0 
& =
 -\sin\Theta_0 \, \sigma_1+\frac{1}{2}\cos\Theta_0\sigma_3 \, ,
\\
A_1 
& = 
-\cos\Theta_0\theta(t)\, \sigma_1 +\dot{\theta}(t)\, \sigma_2 - \frac{1}{2}\sin\Theta_0\theta(t) \, \sigma_3 \, ,
\\
A_2 
& = 
\frac{1}{2}\sin\Theta_0\,\theta^2(t) \, \sigma_1-\frac{1}{4}\cos\Theta_0\,\theta^2(t) \, \sigma_3 
\, .
\end{aligned}
\end{equation}
The time evolution operator for the degenerate subspace is determined by this connection via
\begin{equation}
 \dot{U}(t) = iA_E(t)U(t), \qquad\qquad U(0) = 1.
\end{equation}
Since $A_E(t)$ is constant along the closing segments, we may write for the holonomy $U_\text{f} \equiv U(2\pi+1)=\e^{-i\epsilon\theta(0)\sigma_2} U_\text{c} \, \e^{i\epsilon\theta(0)\sigma_2}$, where $U_\text{c}$ is the holonomy associated to the middle (fluctuating) segment of the curve, \emph{i.e.}, for $0 \leq t <2\pi$. For the latter, we note that $A_0$ in (\ref{WZConnectionMiddleCurve}) is time independent, so that $W(t) \equiv \e^{-iA_0t}U(t)$ satisfies
\begin{equation}
 \dot{W}(t) = iA^I(t)W(t), \qquad\qquad W(0)=1,
\end{equation} 
where 
$A^I(t) = \epsilon A_1^I(t) + \epsilon^2 A_2^I(t)$ 
and 
$A_m^I(t) \equiv \e^{-iA_0t}A_m(t)\e^{iA_0t}$, $m=1,2$,
 are given explicitly by
\begin{equation}\label{conjugatedA1A2}
\begin{aligned}
A_1^I & = \frac{\cos\Theta_0}{\Omega^2}\left[\left(1-\Omega^2-\cos(\Omega t)\right)\theta(t) - \Omega\sin(\Omega t) \dot{\theta}(t)\right]\sigma_1 + \left[\cos(\Omega t)\dot{\theta}(t)-\frac{\sin(\Omega t)}{\Omega}\theta(t)\right]\sigma_2 \\
& \quad + \frac{\sin\Theta_0}{2\Omega^2}\left[\left(4-4\cos(\Omega t)-\Omega^2\right)\theta(t) - 4\sin(\Omega t)\dot{\theta}(t)\right]\sigma_3, \\
A_2^I & = \frac{1}{2}\sin\Theta_0\,\theta^2(t)\sigma_1-\frac{1}{4}\cos\Theta_0\,\theta^2(t)\sigma_3,
\end{aligned}
\end{equation}
where $\Omega = \sqrt{1+3\sin^2\Theta_0}$.  We can integrate this equation perturbatively through a Dyson series up to second order in $\epsilon$, arriving at $W(2\pi)$. Then, $U_\text{c}=e^{i 2\pi A_0}W(2\pi)$, $U_\text{f}$ follows by conjugation of $U_\text{c}$ by the closing segments, and, finally, $U_\text{N}^C=e^{-i\pi}U_\text{f}$ is the holonomy corresponding to our choice of gauge in~(\ref{eigenpm}).  Putting everything together, we get
\begin{equation}
\label{UNCres}
  U_\text{N}(C)  
  = 
  \e^{-i\pi}\e^{-i\epsilon\theta(0)\sigma_2}\e^{i2\pi A_0}
  W(2\pi)
  \e^{i\epsilon\theta(0)\sigma_2}
  \, ,
\end{equation}
where
\begin{align}
 W(2\pi) 
   &=
   \nonumber
    \exp\left(
  i\epsilon \int_0^{2\pi} \de t_1 A_1^I(t_1) 
  + 
  i\epsilon^2 \int_0^{2\pi} \de t_1 A_2^I(t_1)  
   +\frac{\epsilon^2}{2} 
   \int_0^{2\pi} \de t_1 \int_0^{t_1} \de t_2 
  \left[ A_1^I(t_1),\,  A_1^I(t_2) \right] 
   \right)
  \, .
\end{align}
Commuting the second factor in the r.h.s.{} of~(\ref{UNCres}) past the third one, and realizing that $e^{-i\pi} e^{i 2\pi A_0}$ is just the unperturbed holonomy $U_0$ of~(\ref{U0def}), we find
\begin{equation}
\label{UNCU0i}
U_\text{N}(C)=
U_0
\, 
\e^{-i\epsilon\theta(0) \hat{\mathbf{m}} \cdot \boldsymbol{\sigma}} 
W(2\pi)
\e^{i\epsilon\theta(0)\sigma_2}
\equiv U_0 \, U_\epsilon
\, ,
\end{equation}
where the last equation defines the ``noise correction factor'' $U_\epsilon$, and
\begin{equation}
\label{mvdef}
\hat{\mathbf{m}} \cdot \boldsymbol{\sigma}
= -\frac{\cos\Theta_0\sin(2\pi\Omega)}{\Omega} \, \sigma_1
+
\cos(2\pi\Omega) \, \sigma_2
-\frac{2\sin\Theta_0\sin(2\pi\Omega)}{\Omega} \, \sigma_3
\, .
\end{equation}
$U_\text{N}(C)$, as given in~(\ref{UNCres}), (\ref{UNCU0i}), is the holonomy corresponding to a single realization of $\theta(\Phi)$. 
In the following, we will consider an appropriate averaging over all such realizations. Two questions that will have to be answered before attempting that, are, first, which quantity exactly should be averaged and, second, what type of statistics to assume for the fluctuations.
\subsection{Quantifying the effect of noise}
\noindent In the implementation of quantum gates based on non-abelian holonomies,
error quantification relies typically on the concept of fidelity, itself deriving from the Bures metric on density matrices. In our case, that only involves pure states,  the Bures metric reduces to the well-known Fubini-Study metric~\cite{Bengtsson2008}, which is defined on the projective Hilbert space and, hence, ignores phase information. Such a concept of distance is probably too coarse for our purposes, since the non-abelian holonomies we study are generalizations of the abelian Berry's phase, which the Fubini-Study metric cannot detect. We are led then to look for a metric on unitary matrices, rather than states, aiming at a unified geometrical treatment of both the abelian and non-abelian cases. 

\begin{figure}[ht]
\centering
\includegraphics[width=.6 \linewidth]{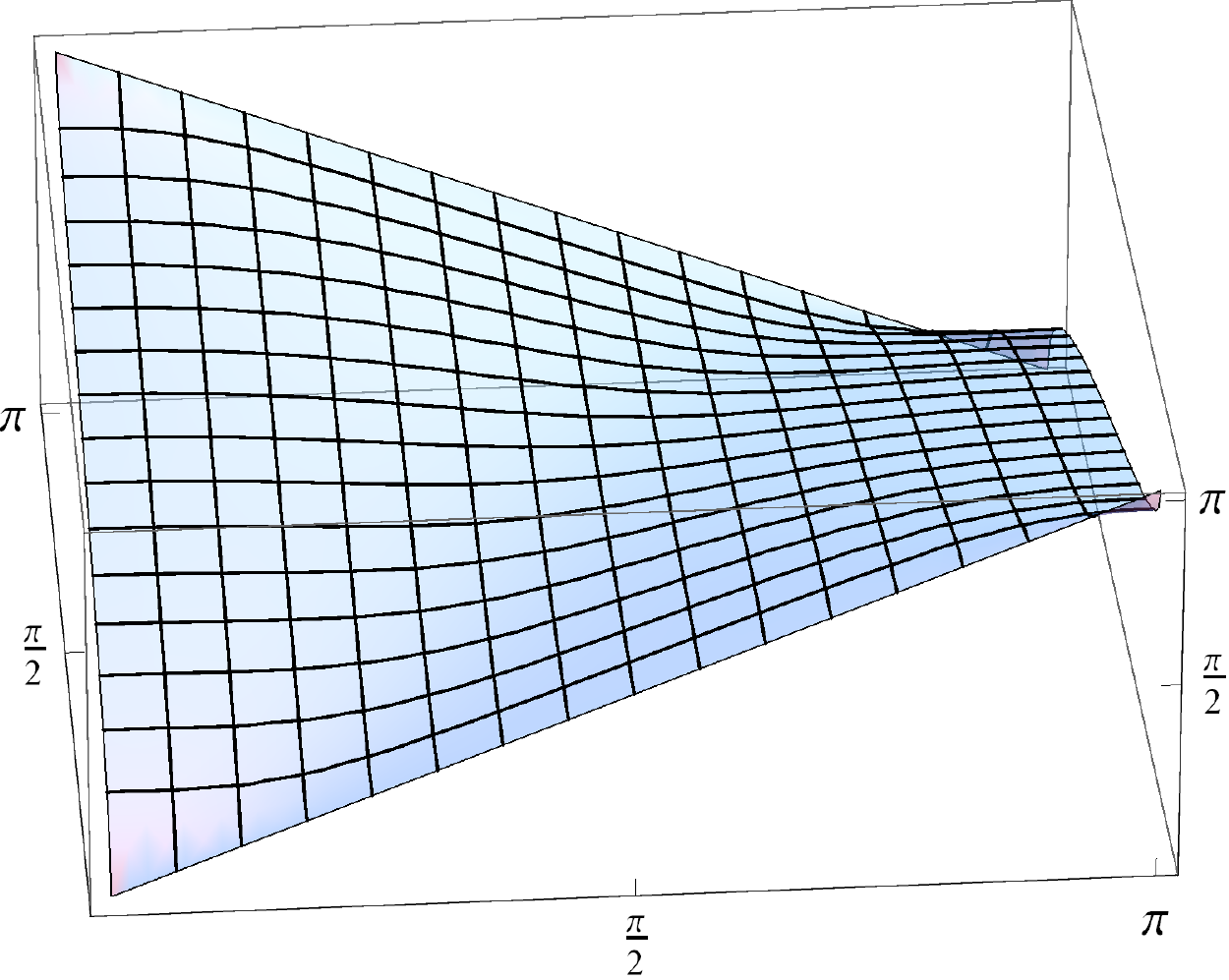}
\caption{Plot of the distance $\tilde{d}_U$ of the $U(2)$ element 
$U=e^{i\alpha}e^{i \theta \hat{\mathbf{n}} \cdot \boldsymbol{\sigma}}$ from the identity, as a function of $\theta$, $\alpha$. For $\alpha=0$ or $\pi$, $\tilde{d}_U$ is linear in $\theta$. For $\theta=\pi/2$, $\tilde{d}_U=\pi/2$, independent of $\alpha$. Finally, for $\theta=\alpha=\pi$, $\tilde{d}_U=0$. This latter behavior reflects the fact that from the south pole $U_\text{S}=e^{i \pi \hat{\mathbf{n}}\cdot \boldsymbol{\sigma}}$ ($\hat{\mathbf{n}}$ arbitrary) of $S^3 \sim SU(2)$, one gets to the identity by ``climbing up'' $\pi$ radians along the $\alpha$-semicircle.}
\label{fig:dPlot}
\end{figure}

For the case of $SU(2)$, a concept of distance between two elements 
\begin{equation}
\label{twoelem}
U=
x_0\mathbf{1} + i\mathbf{x}\cdot\boldsymbol{\sigma}
 \, ,
 \qquad
 V=
y_0\mathbf{1} + i \mathbf{y} \cdot \boldsymbol{\sigma}
 \, ,
\end{equation}
with $x_0^2+|\mathbf{x}|^2=1=y_0^2+|\mathbf{y}|^2$,
is given by
\begin{equation}
\label{def1dist}
d(U,V) 
=
\arccos \left(
\frac{1}{2}\Tr(UV^{\dagger})
\right)
=
 \arccos( x_0 y_0 + \mathbf{x}\cdot\mathbf{y})
\, ,
\end{equation}
which is evidently bi-invariant, and hence, invariant under gauge transformations ---geometrically, it  gives the angle between the points $(x_0,\mathbf{x})$, $(y_0,\mathbf{y})$, representing $U$, $V$, respectively, on $S^3$ (\emph{i.e.}, the underlying manifold of $SU(2)$). Although the holonomies that we compute in the rest of the paper are in fact elements of $SU(2)$, we would like, for completeness, to extend the above definition to the whole of $U(2)$, maintaining its essential property of bi-invariance. The trouble with using~(\ref{def1dist}) for $U(2)$ matrices is that $\Tr UV^\dagger$ is no longer real, in general --- we propose to stay as close as possible to the $SU(2)$ case, by defining
\begin{equation}
\label{def2dist}
\tilde{d}(U,V) 
=
\left| \arccos \left(
\frac{1}{2}\Tr(UV^{\dagger})
\right)
\right|
\, .
\end{equation}
To get some idea about the $\tilde{d}$ distance, consider the general $U(2)$ matrix $U=e^{i\alpha}e^{i \theta \hat{\mathbf{n}}\cdot \boldsymbol{\sigma}}$. Its $SU(2)$ part (the second exponential factor) lies at an angle $\theta$ from the north pole (identity) of $S^3$ ($SU(2)$). From that point, one has still to ``climb up'' an angle $\alpha$ along the $U(1)$ circle, to get to $U$. The distance of this matrix from the identity is $\tilde{d}(U,I) \equiv\tilde{d}_U=| \arccos(e^{i\alpha} \cos\theta)|$  --- we plot it as a function of both $\theta$ and $\alpha$ in Figure~\ref{fig:dPlot}.

In our case, the unitary matrices corresponding to evolution with and without noise are $U_\text{N}(C)=U_0 U_\epsilon$ and $U_0$, respectively (both in $SU(2)$), so that, writing $U_\epsilon=e^{i \epsilon \boldsymbol{\Lambda} \cdot \boldsymbol{\sigma}}$, with $\epsilon \geq 0$, their distance 
comes out equal to
\begin{equation}
\label{distU0UC}
d(U_\text{N}(C),U_0)
=
\arccos \left( \frac{1}{2}\Tr \, U_\epsilon \right) 
=
\epsilon 
|\boldsymbol{\Lambda}|
\equiv 
\epsilon \Lambda
\, ,
\end{equation}
\emph{i.e.}, the effect of adding noise to the precessing magnetic field is to displace the corresponding holonomy by an angle $\epsilon \Lambda$ on $S^3$. The above holds true for each realization of $\theta(\Phi)$, \emph{i.e.}, $\boldsymbol{\Lambda}$ itself  is a vector of stochastic processes. Its explicit form may be deduced from~(\ref{UNCU0i}), 
\begin{equation}
\label{Lambdaexpr1}
U_\epsilon
=
e^{i \epsilon \boldsymbol{\Lambda} \cdot \boldsymbol{\sigma}}
=
\e^{-i\epsilon\theta(0) \hat{\mathbf{m}} \cdot \boldsymbol{\sigma}} 
W(2\pi)
\e^{i\epsilon\theta(0)\sigma_2}
\, ,
\end{equation}
after combining the factors on the r.h.s.{} into a single exponential. Since we are interested in the statistical properties of $\Lambda$, we defer further calculations until after we discuss the statistics of the fluctuations, task to which we now turn.
\subsection{Noise statistics}
\label{sect:StatsofFlucts}
\noindent The function $\theta(\Phi)$ describing curve fluctuations is formally a stochastic process. We do not prescribe any particular statistics for it but only some very general conditions guaranteeing adiabaticity and that no point of the $\Theta_0$ curve is privileged. We denote averaging over all realizations of fluctuations by $\langle\cdot\rangle$. Our minimalist requirements are:
\begin{enumerate}[label=\roman*.]
\item The autocorrelation function depends only on the azimuthal angle difference, $\langle\theta(\Phi_1)\theta(\Phi_2)\rangle=R_{\theta\theta}(\Phi_1-\Phi_2)$, which implies that $\langle\theta^2(\Phi)\rangle = R(0) \equiv \sigma^2$ is a $\Phi$-independent quantity. We also get information about the correlations of the process with its derivatives,
\begin{align}
R_{\theta'\theta}(\Phi_1,\Phi_2)
&\equiv
\left\langle 
\theta'(\Phi_1) \theta(\Phi_2)
\right\rangle 
=
- \left\langle 
\theta(\Phi_1) \theta'(\Phi_2)
\right\rangle 
= R'_{\theta\theta}(\Phi_1-\Phi_2)
\\
R_{\theta'\theta'}(\Phi_1,\Phi_2)
&\equiv
\left \langle 
\theta'(\Phi_1)
\theta'(\Phi_2)
\right \rangle
 = -R''_{\theta\theta}(\Phi_1-\Phi_2)
 \, ,
 \end{align}
where primes denote derivatives w.r.t.{} the argument. Consider now a single, nonzero, mode, $\theta^{(m)}(\Phi)=(2\pi)^{-1/2}(\theta_m\e^{im\Phi}+\theta_m^*\e^{-im\Phi})$ and write $\theta_m = \theta_m^R + i\theta_m^I$. Then, the autocorrelation function is
\begin{equation}
 \begin{aligned}
 R_{\theta \theta}^{(m)}(\Phi_1-\Phi_2)
 & =
  \left \langle 
  \theta^{(m)}(\Phi_1) \theta^{(m)}(\Phi_2) 
 \right \rangle 
 \\
  & = 
  \frac{1}{\pi}\left[
  -2\langle\theta_m^R\theta_m^I\rangle\sin \left( m(\Phi_1+\Phi_2) \right)
  + \langle(\theta_m^R)^2 + (\theta_m^I)^2\rangle \cos \left( m (\Phi_1-\Phi_2) \right) \right. 
  \\
  & \quad \left. + \langle(\theta_m^R)^2 - (\theta_m^I)^2\rangle
  \cos \left( m(\Phi_1+\Phi_2) \right) \right]
  \, ,
\end{aligned}
\end{equation}
from which we deduce that
\begin{equation}
\label{indeqvar}
\left \langle 
\theta_m^R \theta_m^I 
\right \rangle=0
\, ,
\qquad
\left \langle 
(\theta_m^R)^2
\right \rangle
 =
 \left \langle
  (\theta_m^I)^2 \right \rangle
\, ,
\end{equation}
implying
\begin{equation}
\label{autocorr}
R^{(m)}_{\theta\theta}(\Phi_1-\Phi_2)
=
\begin{cases}
\frac{1}{\pi} 
\left \langle 
|\theta_m|^2 
\right \rangle 
\cos \left( m (\Phi_1-\Phi_2) \right)
&
\text{for $m$ nonzero integer}
\\
\frac{1}{2\pi} \left \langle \theta_0^2 \right \rangle
&
\text{for $m=0$}
\end{cases}
\, .
\end{equation}
\item Adiabatic evolution guarantees that a quantum state locks onto an energy level.  We decompose $\theta$ in discrete Fourier components as above. Then, adiabaticity may be enforced by requiring a small amplitude for $\theta$, and by suppressing high frequencies, \emph{e.g.}, by taking $\left \langle|\theta_m|^2 \right \rangle \propto \epsilon m^{-(1+\alpha)}$, with $\epsilon$ appropriately small and $\alpha \geq 0$. Specific choices should take into account that the ``smallness'' mentioned is w.r.t.{} the scale set by the energy difference between the state in question and the closest neighbor connected with it via the time-derivative of the hamiltonian.
\end{enumerate}
\subsection{Noise effects in nuclear quadrupole resonance}
\noindent 
We retake now our calculation, continuing from Eq.~(\ref{Lambdaexpr1}). Our first aim is to get an idea of the size of the effect of noise on the holonomy by calculating the r.m.s.{} value of $\Lambda$. The latter is of order $\epsilon^0$ (with higher order corrections) since we factored out an $\epsilon$ explicitly in, \emph{e.g.},~(\ref{distU0UC}), so, to keep our expressions reasonably complicated, we drop all quadratic and higher terms. This simplifies things considerably, as the leading term in the exponents of all three factors in the r.h.s.{} of~(\ref{Lambdaexpr1}) is of order $\epsilon$, meaning that all the Baker-Campbell-Hausdorff corrections  can be neglected, being at least quadratic in $\epsilon$. Thus, we arrive at
\begin{equation}
\label{dfirstorder}
\boldsymbol{\Lambda} 
 =  
 -\theta(0) \mathbf{m}'  
 +
  \int_0^{2\pi} dt_1 \mathbf{A}_1^I(t_1)
\, ,
\end{equation}
where $\mathbf{m}' \equiv \hat{\mathbf{m}}-\mathbf{e}_2$, and we have written $A_1^I=\mathbf{A}_1^I \cdot \boldsymbol{\sigma}$. The quantity
\begin{equation}
d_\text{rms} 
= 
\sqrt{
\left \langle 
d(U_\text{N}^C,U_0)^2 
\right \rangle}
=
\epsilon
\sqrt{\left \langle \Lambda^2 \right \rangle}
\end{equation}
is a measure of the spread of the distribution of points on the 3-sphere (\emph{i.e.}, the $SU(2)$ manifold) corresponding to all possible evolution operators. Computing $\Lambda^2$ from~(\ref{dfirstorder}) and taking the average over noise realizations, we find
\begin{equation}
\label{AverageSquareDistance}
\begin{aligned}
d_{\text{rms}}^2  
=
\epsilon^2
\left \langle \Lambda^2 \right \rangle
& = 
\epsilon^2
\left[
R_{\theta\theta}(0) ||\mathbf{m}'||^2 
+ 
\int_0^{2\pi} \int_0^{2\pi} \de \Phi_1 \de \Phi_2
\left \langle 
\mathbf{A}_1^I(\Phi_1) \mathbf{A}_1^I(\Phi_2) 
\right \rangle \right. 
\\
& \qquad \left. 
- 2 \mathbf{m}'  \cdot \int_0^{2\pi} \de \Phi_1 
\left \langle 
\mathbf{A}_1^I(\Phi_1) \theta(0) 
\right \rangle 
\right]
\, .
\end{aligned}
\end{equation}
A somewhat lengthy but straightforward computation using (\ref{conjugatedA1A2}), (\ref{mvdef}), and the above mentioned statistical properties of $\theta(\Phi)$, yields
\begin{align}
\label{averageDistance}
d_{\text{rms}}^2 
= 
\epsilon^2
&
\frac{\Omega^{2}-1}{2\Omega^{2}}
\int\limits_0^{2\pi}\!\, \de \sigma R_{\theta\theta}(\sigma) 
\left[ 
\sigma \Omega^{2} 
+ 4 (2\pi-\sigma) 
\left( \Omega^{2}-1\right) \cos(\sigma\Omega)
\right.
\\ \nonumber
& \left. \quad \quad \quad \quad{} -4 \Omega \sin(\sigma\Omega) + 4 \Omega \sin(2\pi\Omega-\sigma\Omega) -4\sigma -2\pi\Omega^{2}+8\pi
\right]
\, .
\end{align}
Note that, to arrive at the above expression, derivatives of $R_{\theta\theta}(\sigma)$ have been integrated by parts --- the extra terms thus produced cancel among themselves.
\subsection{Sample statistics}
\begin{figure}[h]
\centering
\includegraphics[width=1.0 \linewidth]{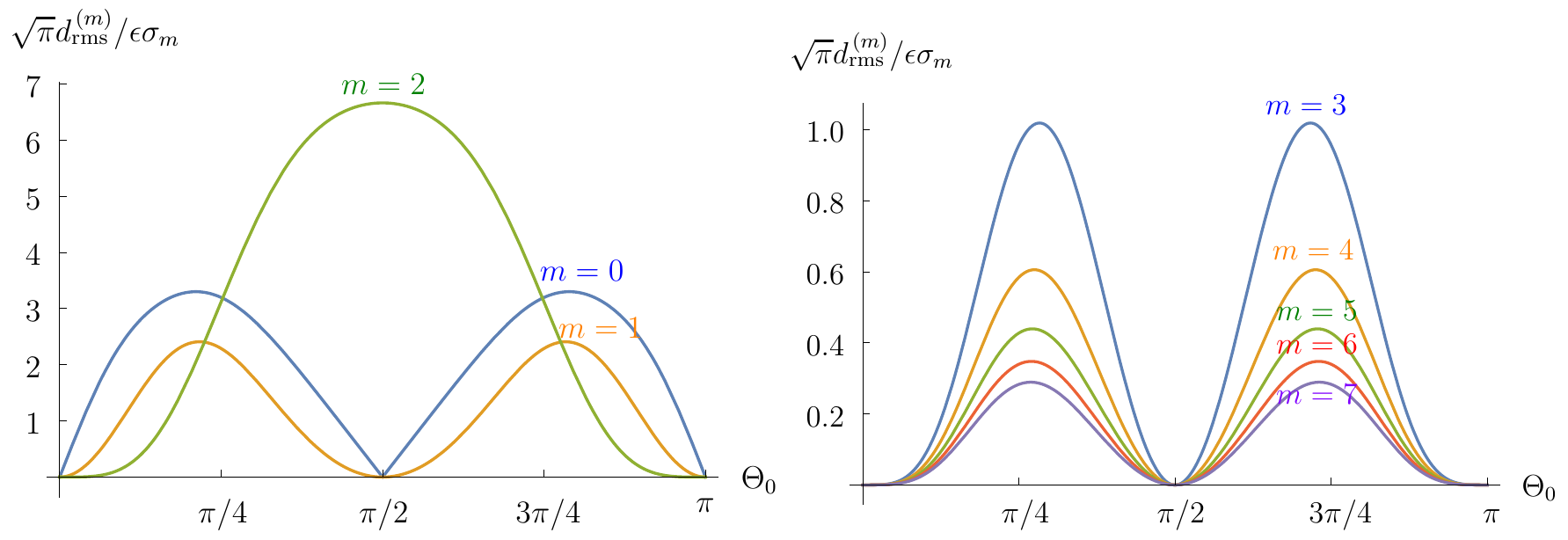}
\caption{Plot of the $d_{\text{rms}}^{(m)}$ vs $\Theta_0$ for $m=0$ (blue), $m=1$ (orange) and $m=2$ (green) (left), and $m=3$ (blue), $m=4$ (orange), $m=5$ (green), $m=6$ (red) and $m=7$ (purple) (right). Note the difference in scale between the two plots.}
\label{fig:m012}
\end{figure}

\noindent 
We consider autocorrelation functions of the form 
\begin{equation}
\label{autocorrspec}
R_{\theta\theta}(\Phi_1-\Phi_2)
=
\begin{cases}
\frac{1}{\pi} 
\sigma_m^2
\cos \left( m (\Phi_1-\Phi_2) \right)
& 
\text{for } m \text{ nonzero integer}
\\
\frac{1}{2\pi} \sigma_0^2
& \text{for } m=0
\end{cases}
\, ,
\end{equation}
corresponding to single-mode noise, with $\sigma_m^2 = \langle|\theta_m|^2\rangle$.
 Then, using (\ref{averageDistance}), the average distance $d_{\text{rms}}^{(m)}$  integrates to
\begin{equation}
\label{eq:dm}
d_{\text{rms}}^{\,(m)}
=
\begin{cases}
\epsilon\frac{\sigma_m}{\sqrt{\pi}}\frac{2\sqrt{m^2+\Omega ^2} \left(1-\Omega^2\right) \sin(\pi  \Omega )}{\Omega |m^2-\Omega^2|}
& 
\text{for } m \text{ nonzero integer}
\\
\epsilon\frac{\sigma_0}{\sqrt{2\pi}}
\frac{\sqrt{\Omega^2-1}}{\Omega^2}\sqrt{4(\Omega^2-1)\sin^2(\pi\Omega)+\pi^2\Omega^2(4-\Omega^2)}
& \text{for } m=0
\end{cases}
\, .
\end{equation}
In Figure~\ref{fig:m012} we plot  $d_{\text{rms}}^{(m)}$ vs $\Theta_0$, for $m=0,1,\ldots,7$. Notice how the $m=2$ case has a completely different behavior, compared to the rest, as well as a different scale. We clarify the origin of this behavior in the following subsection.
\subsection{Understanding the results}
\noindent 
A look at Figure \ref{fig:m012} reveals several characteristics the origin of which is not immediately obvious at an intuitive level (at least not to the authors): all $m$-values, except $m=2$, behave similarly, having common zeros at the poles and the equator, and sharing a more or less common form over the entire range of $\Theta_0$ (the curve $m=0$ differs from the rest in the way it approaches zero at the equator). For $m \geq 3$, the maximum amplitude of $d_\text{rms}^{\, (m)}/\sigma_m$ diminishes with increasing frequency --- a quick check reveals good fit by the law $d_\text{rms}^{\, (m)}|_\text{max} \sim m^{-1}$,  which gets increasingly accurate as $m$ increases.
And then, the $m=2$ curve has a much larger amplitude than all the rest, and reaches its maximum at the equator, where all the other curves have a minimum.

\begin{figure}[h]
\centering
\includegraphics[width=.45 \linewidth]{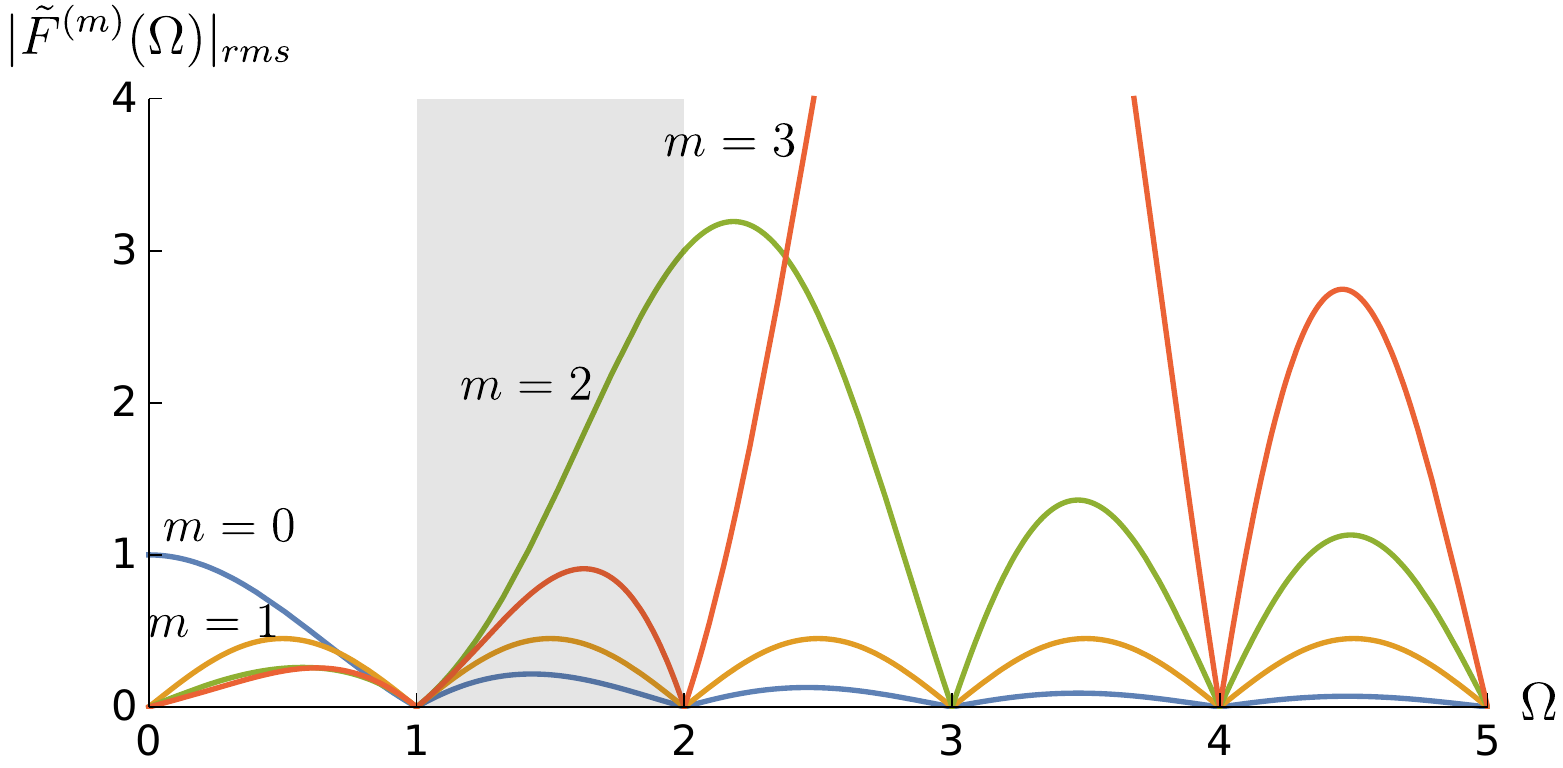}
\hfill
\includegraphics[width=.45 \linewidth]{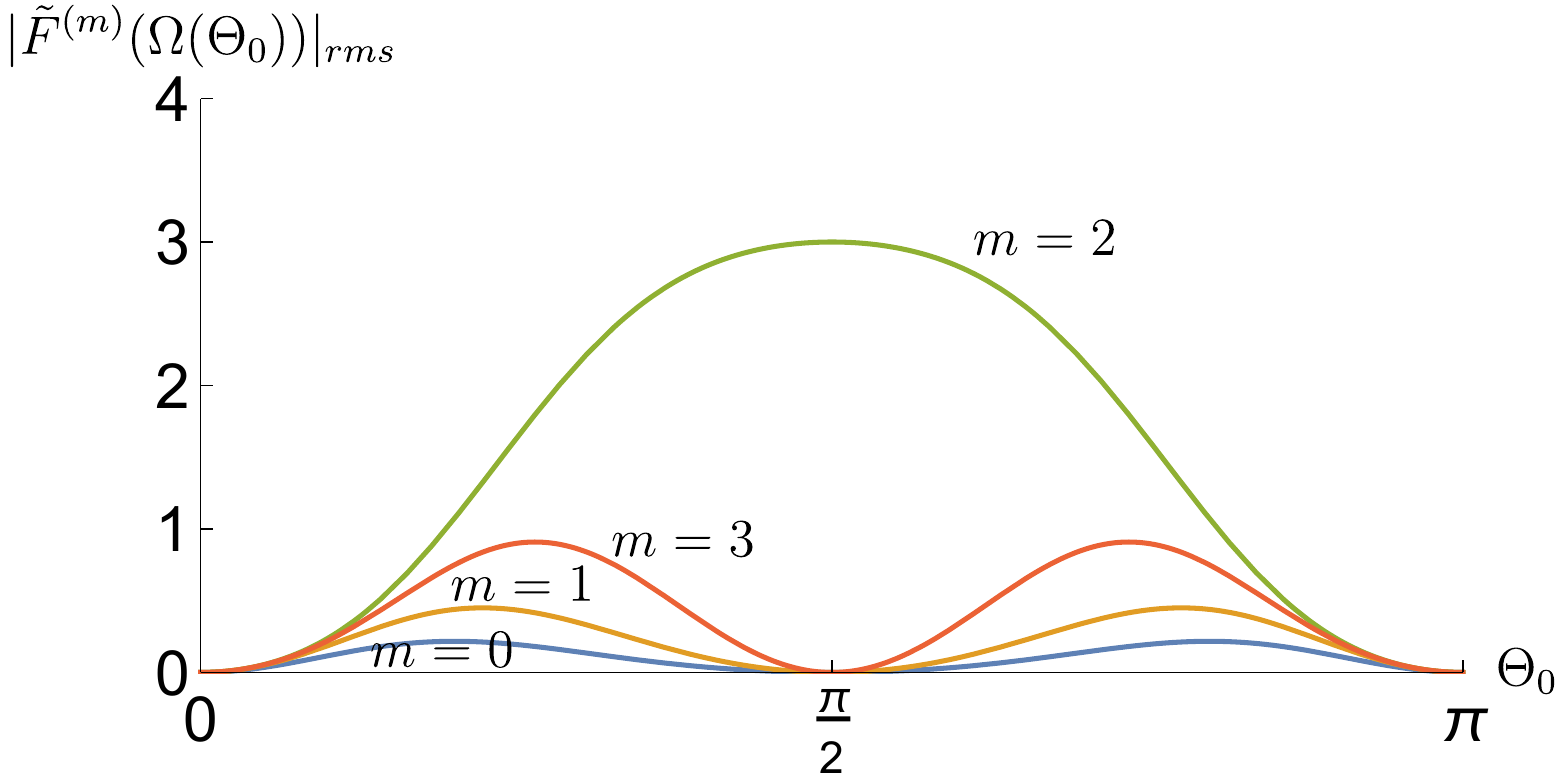}
\caption{Left: Plot of $\sqrt{\langle |\tilde{\mathsf{F}}^{(m)}(\Omega)|^2 \rangle}$, obtained from~(\ref{FT2av}) assuming a  single mode $m$, (\emph{i.e.}, $\sigma_k=\delta_{km}$) for $m=0,1,2,3$. The shaded stripe corresponds to the interval $1 \leq \Omega \leq 2$, over which $\Omega$ ranges (back and forth) as $\Theta_0$ goes from 0 (north pole) to $\pi$ (south pole). Right: Plot of $\sqrt{\langle |\tilde{\mathsf{F}}^{(m)}(\Omega(\Theta_0))|^2 \rangle}$, for the same values of $m$. }
\label{fig:FTm0123}
\end{figure}

We set out to derive these results in a way that makes them obvious, or at least, more predictable.  We begin with the expression for $\boldsymbol{\Lambda}$ in~(\ref{dfirstorder}), and perform two successive $\mathfrak{su}(2)$ basis transformations, motivated by the following considerations: a look at the first of (\ref{WZConnectionMiddleCurve}), shows that $\mathbf{A}_0$ points in some awkward direction in the 13-plane, making an angle $\eta$ with the 3-axis, where 
\begin{equation}
\label{etadef}
\cos \eta= \frac{\cos\Theta_0}{\Omega}
\, ,
\qquad
\sin\eta= \frac{2\sin\Theta_0}{\Omega}
\, .
\end{equation}
Note that subsequent perturbative calculations involve rotating $\mathbf{A}_i$, $i=1,2$ around $\mathbf{A}_0$, as in~(\ref{conjugatedA1A2}), operation that would simplify considerably if $\mathbf{A}_0$ were pointing along, say, the 3-axis. Thus, our first basis transformation, from the original $\{ \sigma_1, \sigma_2, \sigma_3\}$-frame, to a primed one, $\{ \sigma'_1,\sigma'_2,\sigma'_3\}$, consists in a rotation by $\eta$ around the 2-axis, so that $\mathbf{A}_0$ only has a $3'$-component, $\mathbf{A}_0 \cdot \boldsymbol{\sigma}=\frac{\Omega}{2} \sigma'_3$, the new basis vectors being given by
\begin{equation}
\label{firstbasisxation}
\boldsymbol{\sigma}'
=
\left(
\begin{array}{ccc}
\sigma'_1 
 ,
& \sigma'_2 
 ,
& \sigma'_3
\end{array}
\right)
=
\left(
\begin{array}{ccc}
\cos\eta \, \sigma_1 + \sin \eta \, \sigma_3
 ,
&
\sigma_2
 ,
& 
-\sin \eta \, \sigma_1+\cos \eta \, \sigma_3
 \end{array}
 \right)
 \, .
\end{equation}
The above mentioned rotations of $\mathbf{A}_i$ around $\mathbf{A}_0$ mix the $1'$, $2'$-components of $\mathbf{A}_i$ --- their action simplifies if one works instead in the basis $\hat{\boldsymbol{\sigma}}$, given by
\begin{equation}
\label{secondbasisxation}
\hat{\boldsymbol{\sigma}}
=
\left(
\begin{array}{ccc}
\sigma_+
 ,
& \sigma_0
 ,
& \sigma_-
\end{array}
\right)
=
\left(
\begin{array}{ccc}
\sigma'_1+i \sigma'_2
 ,
&
\sigma'_3
 ,
& 
\sigma'_1-i \sigma'_2
 \end{array}
 \right)
 \, ,
\end{equation}
in which the components $A_{i\pm}$ transform by a phase under a $3'$-rotation.
 In this latter basis,
 \begin{equation}
 \label{mptensorb}
 \mathbf{m}' \cdot  \boldsymbol{\sigma}
 =
 -i \left( e^{-i2\pi \Omega}-1 \right) \sigma_+ 
 +
 i \left( e^{i2\pi \Omega}-1 \right) \sigma_-
 \, ,
 \end{equation}
 while
\begin{equation}
\begin{aligned}
A^I_{0} & = \mathbf{A}_0 \cdot \boldsymbol{\sigma}
=
 \frac{\Omega}{2}\sigma_0,\\
A^I_{1} & = \mathbf{A}_1 \cdot \boldsymbol{\sigma}
=
 \frac{1}{\Omega}
\left(
-e^{-i\Omega t}f(t)\sigma_+ 
+ 
\frac{3}{2}\sin\Theta_0\cos\Theta_0 \, \theta(t) \sigma_0 
- e^{i\Omega t} f^{*}(t) \sigma_-
\right)
\, ,
\end{aligned}
\end{equation}
where $f(t) \equiv \theta(t)+i\Omega \dot{\theta}(t)$ is a periodic function, with period $2\pi$. Define now a new (non-periodic) function $\mathsf{f}(t)$ as follows
\begin{equation}
\label{mathsffdef}
\mathsf{f}(t)=f(t)u(t)=\begin{cases}
f(t) & \text{if } 0 \leq t \leq 2\pi
\\
0 & \text{otherwise}
\end{cases}
\, ,
\end{equation}
where $u(t)$ is the unit pulse, equal to 1 for $0 \leq t \leq 2\pi$, and zero elsewhere.
The Fourier transform of $\mathsf{f}(t)$ is given by
\begin{align*}
\label{Fdef}
\tilde{\mathsf{F}}(\omega)
&=
(2 \pi)^{-1/2}\int_{-\infty}^\infty \de t \, \mathsf{f}(t) e^{-i\omega t}
\\
&=
(2 \pi)^{-1/2}\int_0^{2 \pi} \de t \, f(t) e^{-i\omega t}
\, ,
\end{align*}
and is equal to the convolution of the discrete Fourier transform of $f(t)$, with the Fourier transform of $u(t)$. The relevance of these definitions becomes apparent when calculating 
the second term in the r.h.s.{} of~(\ref{dfirstorder}),
\begin{equation}
\label{eq:pdotsigmaNEW}
\int_0^{2\pi} \de t_1 A_1^I(t_1) 
=
\frac{\sqrt{2\pi}}{\Omega}
\left(
-\tilde{\mathsf{F}}(\Omega) \sigma_+ 
+ 
\frac{3\theta_0}{2} \sin\Theta_0 \cos\Theta_0 \, \sigma_0 
- 
\tilde{\mathsf{F}}(\Omega)^* \sigma_-
\right)
\, ,
\end{equation}
where $\theta(t)$ has been expanded in its Fourier modes, as before (see, \emph{e.g.}, the caption of Figure~\ref{fig:fluctuations}). Using~(\ref{dfirstorder}), (\ref{mptensorb}) and (\ref{eq:pdotsigmaNEW}), we may write $\boldsymbol{\Lambda} \cdot \boldsymbol{\sigma}$ in the $\hat{\boldsymbol{\sigma}}$-basis,
\begin{equation}
\label{eq:kThetaGeneral}
 \begin{aligned}
  \boldsymbol{\Lambda} \cdot \boldsymbol{\sigma} & 
  = 
  \left[
  -\frac{\sqrt{2\pi}}{\Omega} \tilde{\mathsf{F}}(\Omega)
  +
  i\theta(0)(e^{-2i \pi \Omega}-1)
  \right]
  \sigma_+ 
  + 
  \frac{3\sqrt{2\pi}\theta_0}{2\Omega} \sin\Theta_0 \cos\Theta_0 \, \sigma_0 
  \\
  &\quad 
  + \left[
  -\frac{\sqrt{2\pi}}{\Omega} \tilde{\mathsf{F}}(\Omega)^*
  -
  i\theta(0)(e^{2i \pi \Omega}-1) 
  \right] 
  \sigma_-
  \, ,
 \end{aligned}
\end{equation}
so that $\Lambda^2=\Lambda_+ \Lambda_-+\Lambda_0^2$ becomes
\begin{equation}
 \Lambda^2 
 = 
 \frac{2\pi}{\Omega^2}
 \big| \tilde{\mathsf{F}}(\Omega) \big|^2 
 + 
 4 \theta(0)^2 \sin^2(\pi\Omega) 
 - 
 \frac{2\sqrt{2\pi}}{\Omega} 
 \theta(0) \, \Im 
 \left[ \left( \e^{i2\pi \Omega}-1 \right) \tilde{\mathsf{F}}(\Omega) 
 \right] 
 + 
 \frac{9\pi\theta_0^2}{2\Omega^2} \sin^2 \Theta_0 \cos^2\Theta_0
 \, ,
\end{equation}
where $\Im [\cdot ]$ denotes imaginary part. For $\tilde{\mathsf{F}}(\omega)$ we find
\begin{equation}
\label{FTomegaexpr1}
\tilde{\mathsf{F}}(\omega)=
\frac{1}{\pi}
e^{-i\pi \omega} \sin \pi \omega
\left(
\frac{\theta_0}{\omega}
+
\sum_{k=1}^\infty
\left(
\frac{k \, \Omega-1}{k-\omega} \theta_k
+
\frac{k \, \Omega +1}{k+\omega}
\theta_k^*
\right)
\right)
\, ,
\end{equation}
so that the average, over the various $\theta(t)$, of its modulus squared, evaluated at $\omega=\Omega$, becomes
\begin{equation}
\label{FT2av}
\langle |\tilde{\mathsf{F}}(\Omega)|^2 \rangle
=
\frac{1}{\pi^2} \sin^2 \pi \Omega
\left(
\frac{\sigma_0^2}{\Omega^2}
+
2
\sum_{k=1}^\infty
\frac{\Omega^2+ k^2 \left( \Omega^4+(k^2-4)\Omega^2+1 \right)}{(k^2-\Omega^2)^2}
\sigma_k^2
\right)
\, .
\end{equation}
Single-mode plots of $|\tilde{\mathsf{F}}(\Omega)|_{\text{rms}}$ appear in Figure~\ref{fig:FTm0123}.
Note that as $\Theta_0$ ranges through the values 0, $\pi/2$, $\pi$, $\Omega$ ranges through 1, 2, and back to 1, respectively. Thus, the only part of the plot on the left in Figure~\ref{fig:FTm0123} relevant to our means  is the shaded stripe, which is plotted again in the same figure on the right, as a function of $\Theta_0$. As expected, $|\tilde{\mathsf{F}}^{(m)}(\Omega)|_{\text{rms}}$ has a maximum close to $\Omega=m$. A curious exception occurs for $m=1$, for which $|\tilde{\mathsf{F}}^{(1)}(\Omega=1)|_{\text{rms}}=0$, but a moment's thought reveals that, in this case, $f(t)=\sqrt{2/\pi} \theta_1^* e^{-it}$, \emph{i.e.}, $f(t)$ has no $e^{it}$-component, so all its energy is at $\omega=-1$. In the interval $[1,2]$, over which we sample $|\tilde{\mathsf{F}}^{(m)}(\Omega)|_{\text{rms}}$, we find the near-maximum of the $m=2$ curve, and the secondary maxima of all the other modes --- this explains the prominence of the $m=2$ curve in Figure~\ref{fig:m012}, as well as several other observed features of $d^{\, (m)}_\text{rms}(\Theta_0)$ that space does not allow us to detail here.
\subsection{Statistics of the Wilczek-Zee holonomy}
\noindent 
For the single-mode $\theta^{(m)}(t)$,
\begin{equation}
\label{thetamdef}
\theta^{(m)}(t)=
\begin{cases}
\frac{1}{\sqrt{2\pi}} (\theta_m e^{imt}+\theta_m^* e^{-imt})
& \text{for $m$ positive integer}
\\
\frac{1}{\sqrt{2\pi}} \theta_0
& 
\text{for $m=0$}
\end{cases}
\, ,
\end{equation} 
we calculate, with the help of Eq.~(\ref{eq:kThetaGeneral}),  the components of $\boldsymbol{\Lambda}^{(m)}$
in the $\boldsymbol{\sigma}'$-frame of Eq.~(\ref{firstbasisxation}),
\begin{equation}
\label{eq:kcomponent}
\boldsymbol{\Lambda}^{(m)} \cdot \boldsymbol{\sigma}
\! 
 =
 \!
\begin{cases} 
 -\frac{4 \left(\Omega ^2-1\right) s_{\pi  \Omega}}{\sqrt{2\pi}\Omega  (m^2-\Omega^2 ) }
 \left(
 (
 \Omega  c_{\pi  \Omega}\theta_m^R+ ms_{\pi  \Omega} \theta_m^I
 )
 \sigma'_1
 +
 (
 \Omega  s_{\pi  \Omega} \theta_m^R- m c_{\pi  \Omega} \theta_m^I
 )
 \sigma'_2
 \right)
 & 
 \text{for $m>0$}
 \\
 \frac{2\theta_0}{\sqrt{2 \pi}\Omega^2}
 \left(
 c_{\pi\Omega} s_{\pi\Omega} (\Omega^2-1)
 \sigma'_1
 +
 s^2_{\pi\Omega}(\Omega^2-1)
 \sigma'_2
 +
 \frac{3 \pi \Omega}{2} s_{\Theta_0} c_{\Theta_0}
 \sigma'_3
 \right)
 &
 \text{for $m=0$}
\end{cases}
\end{equation}
where $\theta_m^R$ and $\theta_m^I$ are the real and imaginary parts of $\theta_m$, and we use the notation $s_x \equiv \sin x$, $c_x \equiv \cos x$. Note that only for $m=0$ does $\boldsymbol{\Lambda}^{(m)}$ acquire a $3'$-component. As a check, one may compute the average modulus squared of $\boldsymbol{\Lambda}^{(m)}$ from these expressions, recovering the results in~(\ref{eq:dm}).

We turn now to the determination of the probability distribution of $\boldsymbol{\Lambda}$, given the one for the Fourier amplitudes $\theta_m$ of the noise $\theta(t)$.
\subsubsection{Distribution of $\boldsymbol{\Lambda}^{(m)}$ for $m \neq 0$}
\label{sec:kthetamneq0}
\begin{figure}[h!]
	\centering
	\includegraphics[width=1.0\linewidth]{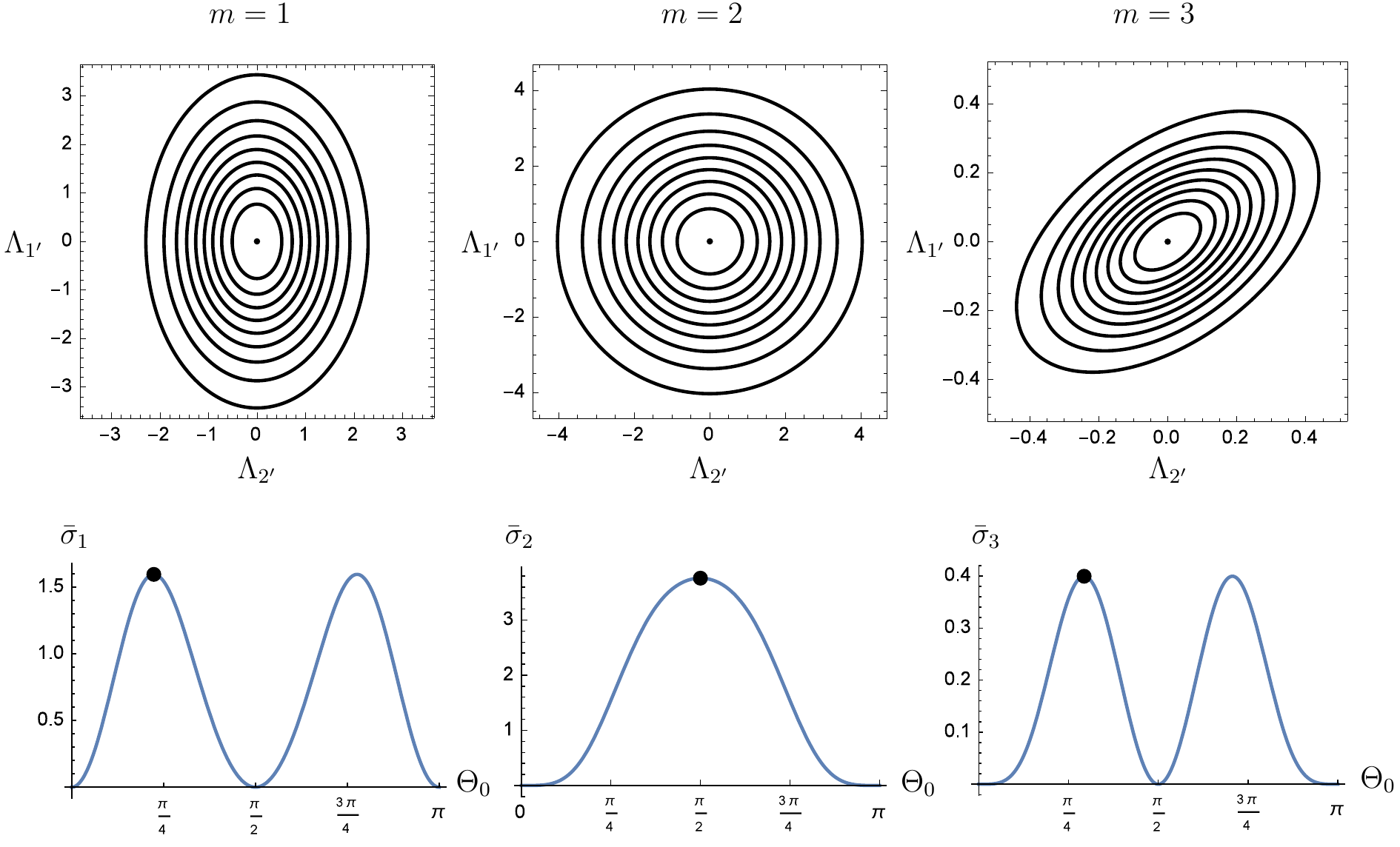}
	\caption{Top row: contour plots of the probability density of  $\boldsymbol{\Lambda}^{(m)}$, for $m=1,2,3$, in the $1'$-$2'$ plane. The plots are 2D since the $3'$-component of $\boldsymbol{\Lambda}^{(m)}$ vanishes for $m \neq 0$. Bottom row:  plots of the standard deviation $\bar{\sigma}_m$ \emph{vs.{}} $\Theta_0$ --- the dots indicate the values of $\Theta_0$ used for the corresponding top row plots. The overall scale of the plots is determined by taking $\langle|\theta_m|^2\rangle\propto m^{-1}$.}
	\label{fig:contourPlots}
\end{figure}

\begin{figure}[h!]
\raisebox{3ex}{\includegraphics[width=.53\linewidth]{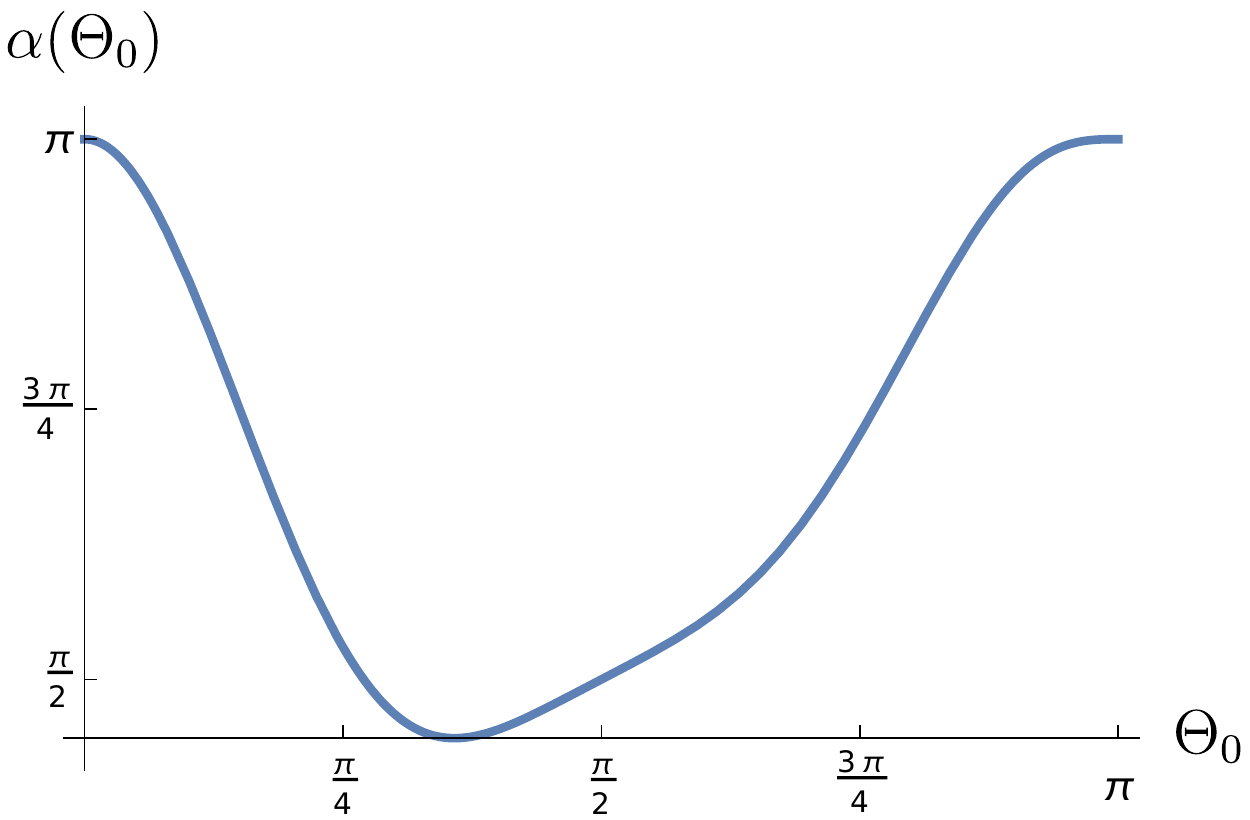}}
\hfill
\raisebox{0ex}{\includegraphics[width=.4\linewidth]{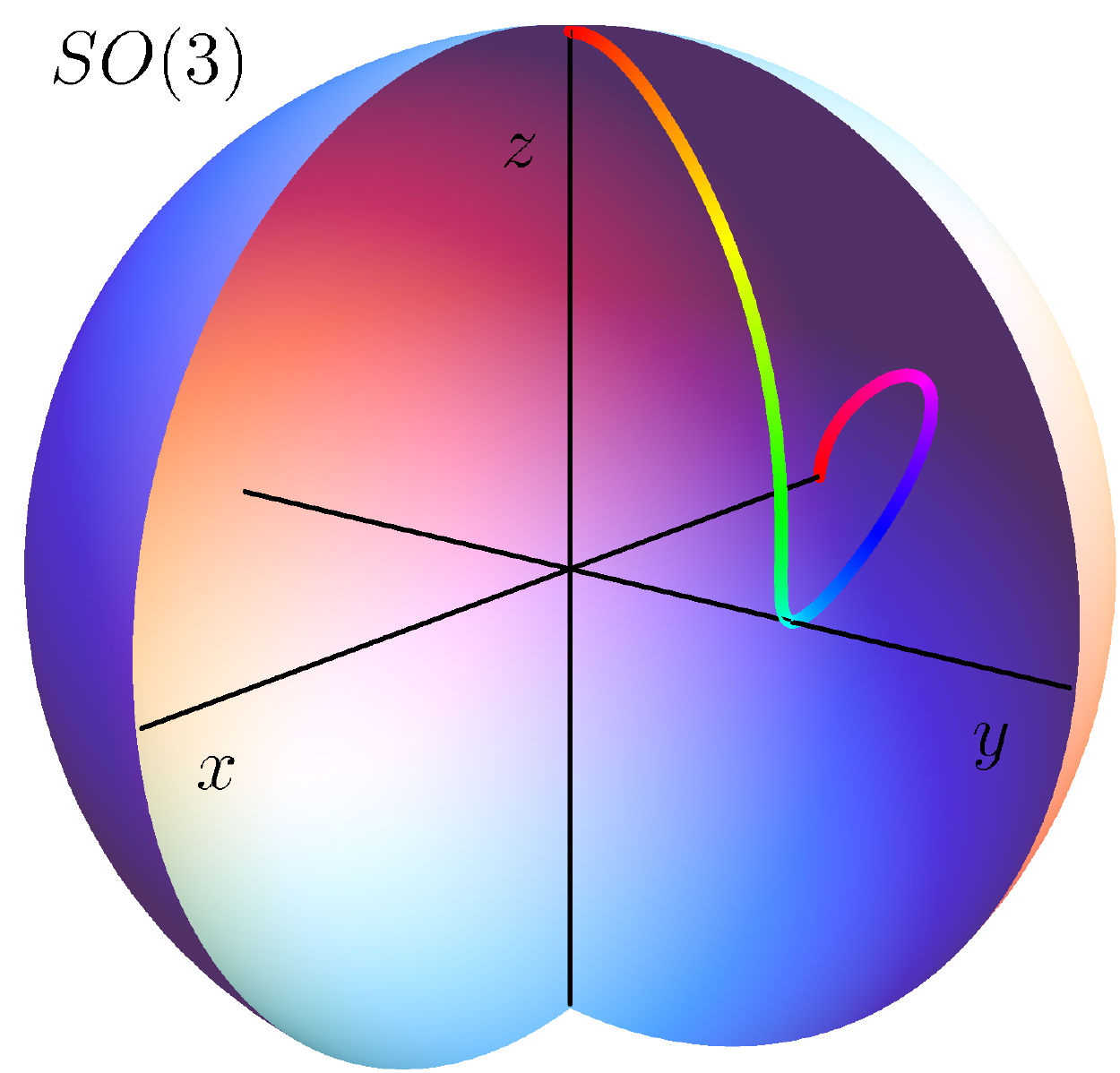}}
\caption{Left: plot of the rotation angle $\alpha(\Theta_0)$ of $R(\Theta_0)$, for $0 \leq \Theta_0 \leq \pi$. Right: plot of the path of $R(\Theta_0)$ in $SO(3)$, in the angle-axis parametrization, for $0 \leq \Theta_0 \leq \pi$. The curve starts  at $(0,0,\pi)$ for $\Theta_0=0$, passes through $(0,\pi/2,0)$ for $\Theta_0=\pi/2$, and ends at $(-\pi,0,0)$ for $\Theta_0=\pi$. }
\label{fig:RtotalPlot}
\end{figure}

\noindent Considering random Fourier amplitudes $\theta_m$, obeying a normal distribution with zero mean and width that scales appropriately with frequency, suffices to account for physically reasonable fluctuations complying with adiabaticity. For $m\neq 0$ the real and imaginary components of the Fourier amplitude $\theta_m$ must be statistically independent and have the same standard deviation (see~(\ref{indeqvar})). Suppose that the probability distributions of both $\theta_m^R$ and $\theta_m^I$ are given by
\begin{equation}
\label{eq:distribTheta}
P\left(\theta_m^{R,I}=\theta\right)
=
\frac{1}{\sqrt{2\pi} \sigma_m} \e^{-\frac{\theta^2}{2\sigma_m^2}}
\, .
\end{equation}
Putting 
$\boldsymbol{\Lambda}^{(m)}
=
(\Lambda_{1'},\Lambda_{2'},\Lambda_{3'})$, 
the first of~(\ref{eq:kcomponent}) implies that
\begin{align}
\left(
\begin{array}{c}
\theta_m^R
\\
\theta_m^I
\end{array}
\right)
&=
-\frac{\sqrt{2\pi} (m^2-\Omega^2 )}{4m s_{\pi\Omega} \left( \Omega^2-1 \right) }
\left(
\begin{array}{cc}
m c_{\pi \Omega}
&
m s_{\pi \Omega}
\\
\Omega s_{\pi \Omega}
&
-\Omega c_{\pi \Omega}
\end{array}
\right)
\left(
\begin{array}{c}
\Lambda_{1'}
\\
\Lambda_{2'}
\end{array}
\right)
\, ,
\end{align}
and $\Lambda_{3'}=0$, which, in turn, leads to
\begin{equation}
\label{eq:pkxkyDistrib}
P 
\left( 
\boldsymbol{\Lambda}^{(m)}
=
(\Lambda_{1'},\Lambda_{2'},\Lambda_{3'})
\right)
=
\frac{m \Omega}{2\pi\bar{\sigma}^2_m}
\e^{-\frac{1}{2\bar{\sigma}_m^2}
(m^2 (c_{\pi\Omega} \Lambda_{1'} + s_{\pi\Omega} \Lambda_{2'})^2
+ \Omega^2 
(-s_{\pi\Omega} \Lambda_{1'} + c_{\pi\Omega} \Lambda_{2'})^2)}\delta(\Lambda_{3'})
\, ,
\end{equation}
where
\begin{equation}
\label{eq:s1s2}
\bar{\sigma}_m
=
\frac{4m s_{\pi\Omega} (\Omega^2-1)}{{\sqrt{2\pi}(m^2-\Omega^2)}} \sigma_m
\, .
\end{equation}
Figure \ref{fig:contourPlots} shows equiprobability curves on the $\Lambda_ {3'}=0$ plane for single-mode fluctuations with frequencies $m=1,2,3$. The angle $\Theta_0$ of the unperturbed curve has been chosen so as to maximize the standard deviation $\tilde{\sigma}_m$ of Eq.~(\ref{eq:s1s2}). All these curves are ellipses (see the exponent in~(\ref{eq:pkxkyDistrib})), with their axes rotated by an angle $\pi \Omega$ w.r.t.{} the $1'$-$2'$ frame. As $\Theta_0$ varies, the ellipses rotate and, at the same time,  expand or stretch, as described by~(\ref{eq:s1s2}).
To get a better idea of the orientation of these ellipses, we define a further rotation in $\mathfrak{su}(2)$, around the $3'$-axis by an angle $\pi \Omega$, taking the primed frame $\boldsymbol{\sigma}'$ to the tilded one $\tilde{\boldsymbol{\sigma}}$, under which the exponent in the r.h.s.{} of~(\ref{eq:pkxkyDistrib}) becomes proportional to $m^2 \Lambda_{\tilde{1}}^2+\Omega^2 \Lambda_{\tilde{2}}^2$, in other words, the $\tilde{1}$, $\tilde{2}$-axes coincide with the axes of the ellipses shown in Figure~\ref{fig:contourPlots}. Combining this rotation with the one taking $\boldsymbol{\sigma}$ to $\boldsymbol{\sigma}'$, (see~(\ref{firstbasisxation})), we find
the composite rotation $R(\Theta_0)$ that takes the $\boldsymbol{\sigma}$-coordinates of $\boldsymbol{\Lambda}$ to the  $\tilde{\boldsymbol{\sigma}}$-ones,
\begin{equation}
\label{totalR}
\left(
\begin{array}{c}
\Lambda_{\tilde{1}}
\\
\Lambda_{\tilde{2}}
\\
\Lambda_{\tilde{3}}
\end{array}
\right)
=
\frac{1}{\Omega}
\left(
\begin{array}{ccc}
c_{\Theta_0} c_{\pi \Omega}
&
\Omega s_{\pi \Omega}
&
2 s_{\Theta_0} c_{\pi \Omega}
\\
-c_{\Theta_0} s_{\pi \Omega}
&
\Omega c_{\pi \Omega}
&
-2 s_{\Theta_0} s_{\pi \Omega}
\\
-2 s_{\Theta_0}
&
0
&
c_{\Theta_0}
\end{array}
\right)
\left(
\begin{array}{c}
\Lambda_{1}
\\
\Lambda_{2}
\\
\Lambda_{3}
\end{array}
\right)
\, .
\end{equation}
Figure~\ref{fig:RtotalPlot} shows a plot of the corresponding rotation angle $\alpha(\Theta_0)$ and the path of $R(\Theta_0)$ in $SO(3)$, for $0 \leq \Theta_0 \leq \pi$.
Note that if the value taken by $\theta_m$ in a particular realization of $\theta(t)$ is real, the corresponding $\boldsymbol{\Lambda}^{(m)}$ is along the $\tilde{1}$-direction, while if $\theta_m$ turns out purely imaginary, $\boldsymbol{\Lambda}^{(m)}$ points along the direction $\tilde{2}$.
\subsubsection{Distribution of $\boldsymbol{\Lambda}^{(0)}$}
\begin{figure}[h!]
\raisebox{0ex}{\includegraphics[width=.65\linewidth]{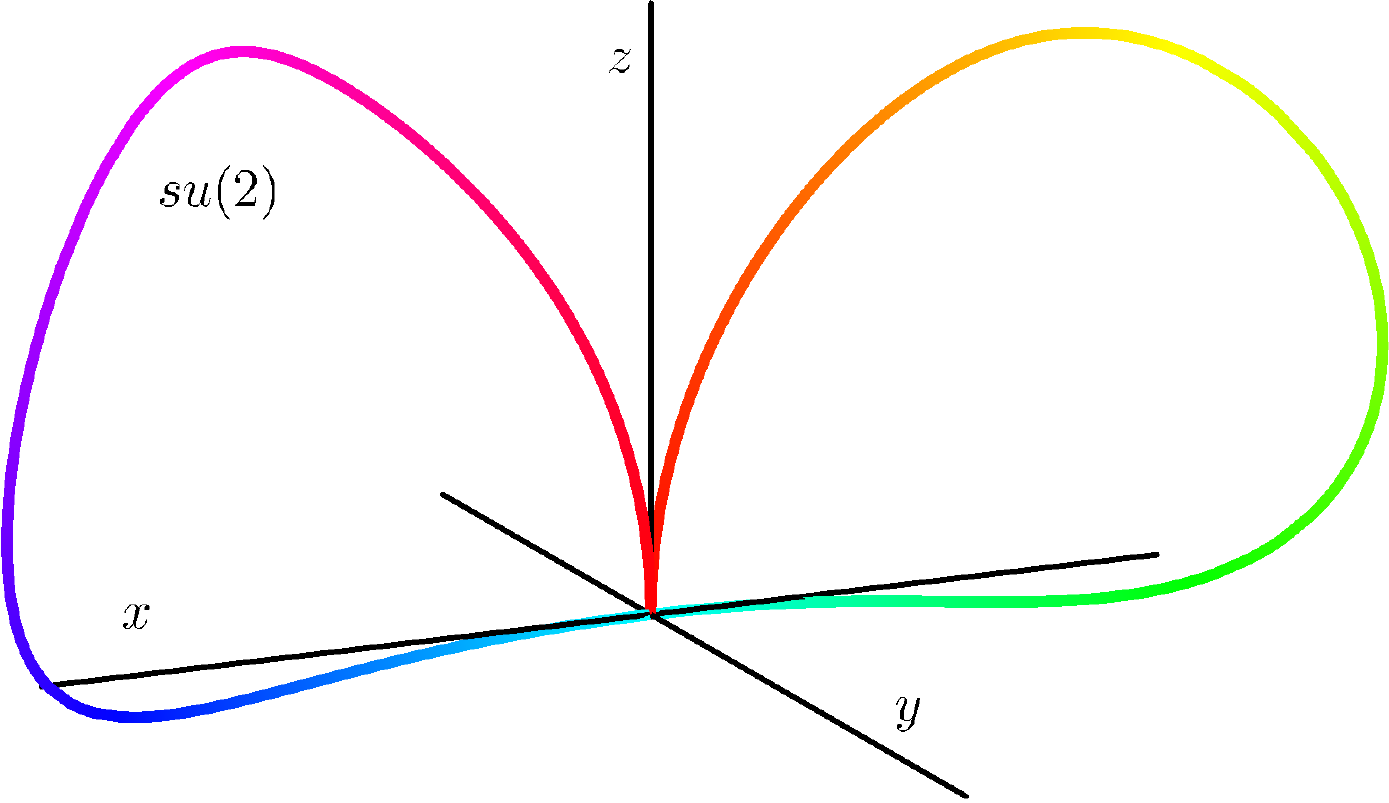}}
\caption{Plot of $\mathbf{n}$ (see Eq.~(\ref{ndef})) \emph{vs.} $\Theta_0$, for $0 \leq \Theta_0 \leq \pi$, in the $\boldsymbol{\sigma}$-frame. The curve starts at the origin of $\mathfrak{su}(2)$ for $\Theta_0=0$, pointing upwards, and traces a loop with $x>0$, returning to the origin for $\Theta_0=\pi/2$. It then traces a second loop, for $ \pi/2 \leq \Theta \leq \pi$, obtained from the first one by reflection in the $y$-$z$ plane.}
\label{fig:Lambda0Plot}
\end{figure}

\noindent 
The second of~(\ref{eq:kcomponent}) implies that $\boldsymbol{\Lambda}^{(0)}=\theta_0 \mathbf{n}$, where the $\Theta_0$-dependent $\mathbf{n}$ is given by
\begin{align}
\mathbf{n} \cdot \boldsymbol{\sigma}
&=
\frac{2}{\sqrt{2\pi}\Omega^2}
\left(
(\Omega^2-1) s_{\pi \Omega} \,  \sigma_{\tilde{1}}
+
\frac{3\pi \Omega}{2}  c_{\Theta_0} s_{\Theta_0} \, \sigma_{\tilde{3}}
\right)
\nonumber
\\
&=
\frac{2}{\sqrt{2\pi}\Omega^2}
\left(
(\Omega^2-1)c_{\pi \Omega} s_{\pi \Omega} \,  \sigma_{1'}
+ 
(\Omega^2-1) s_{\pi \Omega}^2 \,  \sigma_{2'}
+
\frac{3 \pi \Omega}{2}  c_{\Theta_0} s_{\Theta_0} \, \sigma_{3'}
\right)
\label{ndef}
\\
&=
\frac{1}{\sqrt{8\pi} \Omega^3}
\lbrace
2c_{\Theta_0} 
[
-3\pi \Omega ( 1-c_{2\Theta_0} ) + (\Omega^2-1) s_{2\pi\Omega} 
]
\sigma_1
+
4\Omega (\Omega^2-1) s^2_{\pi \Omega}\, 
\sigma_2
\nonumber
\\
 & \qquad \qquad \qquad {}+
s_{\Theta_0}
[
3\pi \Omega(1+c_{2\Theta_0}) + 4(\Omega^2-1) s_{2\pi \Omega}
] \,
\sigma_3
\rbrace
\nonumber
\, ,
\end{align}
the first equality above making it evident that $\mathbf{n}$ is orthogonal to the $\tilde{2}$ direction ---  a plot of $\mathbf{n}$ \emph{vs.} $\Theta_0$ appears in Figure~\ref{fig:Lambda0Plot}.
$\boldsymbol{\Lambda}^{(0)}$ lies along the line spanned by $\mathbf{n}$, with probability density that of $\theta_0$, scaled by  
$|\mathbf{n}|
=
2/(\sqrt{2\pi}\Omega^2)
(9\pi^2 \Omega^2 s^2_{2\Theta_0}/16+(\Omega^2-1)^2 s^2_{\pi \Omega})^{1/2}
$.
In particular, if $\theta_0$ follows a normal distribution, with zero average and standard deviation $\sigma_0$, $\boldsymbol{\Lambda}^{(0)}$ will also follow a normal distribution along $\mathbf{n}$, with zero average and standard deviation $|\mathbf{n}| \sigma_0$. In terms of its $\tilde{\boldsymbol{\sigma}}$-components, the probability density of $\boldsymbol{\Lambda}^{(0)}$ is given by
\begin{equation}
P \left(
\boldsymbol{\Lambda}^{(0)}
=
( \Lambda_{\tilde{1}}, \Lambda_{\tilde{2}},\Lambda_{\tilde{3}} ) 
\right)
=
\delta \left( \Lambda_{\tilde{1}}   -\mu \Lambda_{\tilde{3}}   \right)
\delta(\Lambda_{\tilde{2}})
\frac{1}{\sqrt{2\pi}\bar{\sigma}_0}
e^{-\frac{\Lambda_{\tilde{3}}^2}{2\bar{\sigma}_0^2}}
\, ,
\end{equation}
where
\begin{equation}
\label{eq:sigma0muDefs}
\bar{\sigma}_0
=
\frac{3\pi}{\sqrt{2\pi}\Omega}  c_{\Theta_0} s_{\Theta_0} \sigma_0
\, ,
\qquad
\mu
=
\frac{2 (\Omega^2-1) s_{\pi\Omega} }{3\pi \, \Omega \,  c_{\Theta_0} s_{\Theta_0}}
\, .
\end{equation}
\subsubsection{Distribution of $\boldsymbol{\Lambda}$: the general case}
\noindent 
We consider now the general case of a multi-mode noise $\theta(t)$. Since $\boldsymbol{\Lambda}$ is linear in $\theta(t)$, we obtain
\begin{equation}
\label{eq:kthetageneral}
\boldsymbol{\Lambda}=\boldsymbol{\Lambda}^{(0)}
+
\sum_{m=1}^{\infty}
\boldsymbol{\Lambda}^{(m)}
\, ,
\end{equation}
where, in the tilded frame,
\begin{equation}
\label{eq:kcomponenttilded}
\boldsymbol{\Lambda}^{(m)}
\! 
 =
 \!
\begin{cases} 
 -\frac{4 \left(\Omega ^2-1\right) s_{\pi  \Omega}}{\sqrt{2\pi}\Omega  (m^2-\Omega^2 ) }
 \left(
\Omega  \theta_m^R \, 
 \sigma_{\tilde{1}}
 -
 m \theta_m^I \, 
 \sigma_{\tilde{2}}
 \right)
 & 
 \text{for $m$ positive integer}
 \\
 \frac{2\theta_0}{\sqrt{2 \pi}\Omega^2}
 \left(
(\Omega^2-1) s_{\pi\Omega} 
\, 
 \sigma_{\tilde{1}}
 +
 \frac{3 \pi \Omega}{2} c_{\Theta_0} s_{\Theta_0}
 \, 
 \sigma_{\tilde{3}}
 \right)
 &
 \text{for $m=0$}
\end{cases}
\, .
\end{equation}
For the second term in the r.h.s.{} of~(\ref{eq:kthetageneral}) the statistical independence of the various $\boldsymbol{\Lambda}^{(m)}$ implies
\begin{equation}
 P 
 \left( 
 \sum_{m=1}^{\infty} \boldsymbol{\Lambda}^{(m)}
 =
 ( \Lambda_{\tilde{1}}, \Lambda_{\tilde{2}},\Lambda_{\tilde{3}} )
 \right)
 =
 \frac{1}{2\pi \alpha_1 \alpha_2} 
 \e^{-\frac{1}{2} \left( 
 \frac{\Lambda_{\tilde{1}}^2}{\alpha_1^2}
 +
\frac{\Lambda_{\tilde{2}}^2}{\alpha_2^2} 
\right)
}
 \delta(\Lambda_{\tilde{3}})
 \, , 
\end{equation}
where
\begin{equation}
\label{eq:alphaDefs}
\alpha_1
=
\sqrt{\sum_{m=1}^\infty  \frac{\bar{\sigma}_m^2}{m^2}} 
\, ,
\qquad \qquad
\alpha_2
=
\frac{1}{\Omega}\sqrt{\sum_{m=1}^\infty  \bar{\sigma}_m^2}
\, .
\end{equation}
Convolution of the above expression with the probability distribution of $\boldsymbol{\Lambda}^{(0)}$ gives finally
\begin{equation}
\label{eq:kGeneralDist}
\begin{aligned}
P \left( 
\boldsymbol{\Lambda}
=
 ( \Lambda_{\tilde{1}}, \Lambda_{\tilde{2}},\Lambda_{\tilde{3}} )
\right)
&=
\frac{1}{(2\pi)^{3/2}\alpha_1\alpha_2\bar{\sigma}_0}
\e^{-\frac{1}{2}
\left[
\frac{(\Lambda_{\tilde{1}}- \mu \Lambda_{\tilde{3}})^2}{\alpha_1^2}
+
\frac{\Lambda_{\tilde{2}}^2}{\alpha_2^2}
+
\frac{\Lambda_{\tilde{3}}^2}{\bar{\sigma}_0^2}
\right]
}
\, .
\end{aligned}
\end{equation}
Equiprobability surfaces for $\boldsymbol{\Lambda}$ are ellipsoids with one axis along $\tilde{2}$, and the other two rotated in the $\tilde{1}$-$\tilde{3}$ plane. For each $\Lambda_{\tilde{3}} = \text{constant}$ cross-section we get a planar ellipse with semiaxes $\alpha_1$, $\alpha_2$,  shifted by  $\mu \Lambda_{\tilde{3}}$ along the $\tilde{1}$-axis.
\begin{figure}[h!]
	\centering
	\includegraphics[width=1.0\linewidth]{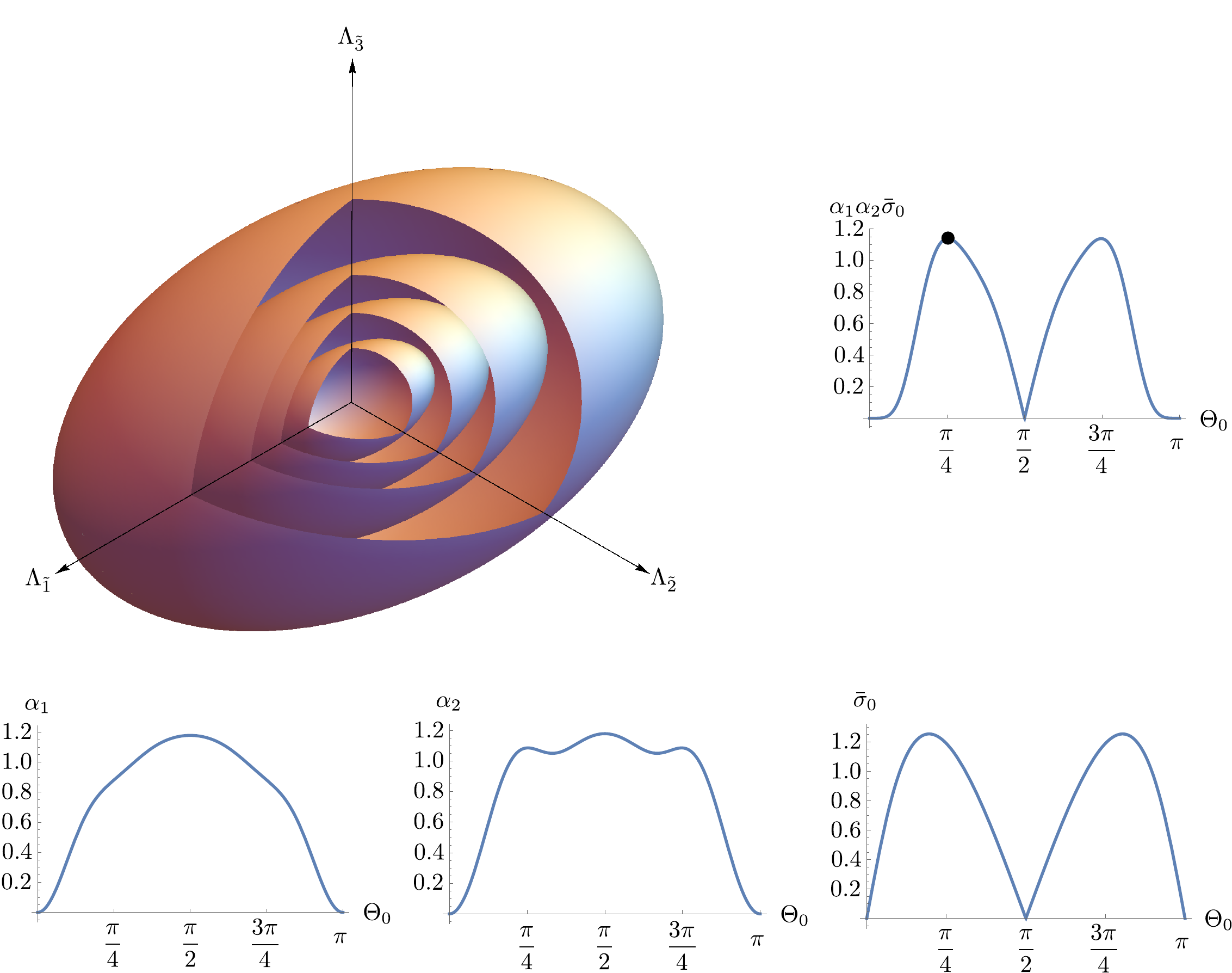}
	\caption{Top row, left: Surfaces of constant probability density for $\boldsymbol{\Lambda}$, in the $\tilde{\boldsymbol{\sigma}}$-frame, assuming the standard deviation $\sigma_m$ of the mode with frequency $m$  proportional to $m^{-1}$.	Top row, right: The product of the three semiaxes of the gaussian probability distribution of $\boldsymbol{\Lambda}$, as a function of the precession angle $\Theta_0$ --- the value of $\Theta_0$ used for the plot on the left is the one that maximizes this product, denoted by a dot in the plot. Bottom row: Plot of the variation of the individual semiaxes with $\Theta_0$. }
\label{fig:equiprob}
\end{figure}
\section{CONCLUDING REMARKS}
\label{conclusions}
\noindent 
We studied the effect of noise on the Wilczek-Zee holonomy of the nuclear quadrupole resonance hamiltonian. An arbitrary periodic noise was decomposed into its Fourier components, and the effect of individual frequencies was analyzed. We found that all frequencies $m \neq 2$ behave similarly: their effect tends to vanish when the unperturbed precession takes place close to the poles or the equator (\emph{i.e.}, when  $\Theta_0$ approaches 0, $\pi/2$, or $\pi$), while it attains its maxima at two intermediate angles, close to $\pi/4$ and $3\pi/4$, respectively. This behavior, as function of $\Theta_0$, closely resembles the abelian case, where the geometric phase accumulated during precession is proportional to the enclosed solid angle, so that, close to the equator, the effect of noise on the solid angle cancels out. If all frequencies produced, more or less,  this same effect, then the experimental physicist seeking to minimize the effects of noise would have a clear-cut solution available: make the unperturbed magnetic field precess close to the equator. But this simple scenario is upset by the presence of the $m=2$ harmonic, which attains its maximum exactly at the equator, and, moreover, affects the holonomy considerably more than the other frequencies, \emph{e.g.}, about seven times more than the $m=3$ component, assuming equal amplitudes. Thus, for a specific noise spectrum, the precession angle(s) $\Theta_0$ at which noise effects get minimized can very well be away from the equator. This predominance of the $m=2$ harmonic seems to be reported here for the first time, and may be considered as a signature of the truly non-abelian nature of the system studied. 

It would have been interesting to be able to compare our results with those in~\cite{solinas2004robustness,solinas2012stability}. However, our analysis in only first-order in the noise amplitude, while the use of fidelity as indicator of gate performance entails a quadratic effect. Thus, a meaningful comparison can only be attempted if the logarithm  $\boldsymbol{\Lambda}$ of the unitary correction to the holonomy is known to second order in the noise amplitude, and we have advanced enough in this calculation to know that the results are much more complicated than the first-order ones reported here.
On the other hand, it should be relatively easy to repeat the numerical study using sinusoidal noise, aiming, for example, to verify the exceptional role of the $m=2$ component, and the consequent departure from the abelian behavior. Another question that could be settled this way would be whether the (partial) failure of adiabaticity reported 
in~\cite{solinas2004robustness} persists when the smoother sinusoidal noise is used. 

On the analytical front, it would be interesting to consider \emph{quantum} noise, \emph{i.e.}, replace the classical magnetic field $\mathbf{B}(t)$  with a quantum vector operator $\mathbf{A}$, the quantum fluctuations of which would play the role of noise. In the abelian case, such a treatment gives rise to a novel effect: part of the correction to the geometric phase involves the commutator of certain components of $\mathbf{A}$~\cite{Agu.Chr.Guz:15a}. A different (and earlier) approach,  considers a quantized field, and predicts a vacuum-induced geometric phase (see~\cite{Fue.Car.Bos.Ved:02} for the original theoretical treatment, \cite{Liu.Fen.Wan:11,Lar:12,Wan.Wei.Lia:15} for subsequent theoretical debate, and the recent experimental observation reported in~\cite{Gas.Ber.Abd.Pec.Fil.Wal:16}) --- we wonder what the corresponding statements, for either approach, would be in the non-abelian case. Another line of research, that we currently pursue, is a geometric description of the Wilczek-Zee holonomy, based on the recent generalization of the Majorana stellar representation to the case of multipartite antisymmetric states~\cite{Chr.Guz.Han.Ser:21}. The subset of the latter that are $\wedge$-factorizable, \emph{i.e.}, that can be written in the form of a Slater determinant, are in 1-to-1 correspondence with linear subspaces of the Hilbert space, of the type employed, as degenerate subspaces of a Hamiltonian,  in the Wilczek-Zee formalism. Cyclic evolution of the Hamiltonian results in the degenerate subspace tracing out a closed curve in the corresponding Grassmannian, which, in principle, completely determines the Wilczek-Zee holonomy. It would be very interesting to see if rotational symmetries of the degenerate subspaces lead to robust holonomies, as it has been shown recently to happen in the abelian case~\cite{Agu.Chr.Guz.Han.Ser:20}.

\section*{Data Availability}
The data that support the findings of this study are available from the corresponding author
upon reasonable request.
\bibliographystyle{apsrev4-1}
\bibliography{WZstochastic}

\begin{thebibliography}{64}%
\makeatletter
\providecommand \@ifxundefined [1]{%
 \@ifx{#1\undefined}
}%
\providecommand \@ifnum [1]{%
 \ifnum #1\expandafter \@firstoftwo
 \else \expandafter \@secondoftwo
 \fi
}%
\providecommand \@ifx [1]{%
 \ifx #1\expandafter \@firstoftwo
 \else \expandafter \@secondoftwo
 \fi
}%
\providecommand \natexlab [1]{#1}%
\providecommand \enquote  [1]{``#1''}%
\providecommand \bibnamefont  [1]{#1}%
\providecommand \bibfnamefont [1]{#1}%
\providecommand \citenamefont [1]{#1}%
\providecommand \href@noop [0]{\@secondoftwo}%
\providecommand \href [0]{\begingroup \@sanitize@url \@href}%
\providecommand \@href[1]{\@@startlink{#1}\@@href}%
\providecommand \@@href[1]{\endgroup#1\@@endlink}%
\providecommand \@sanitize@url [0]{\catcode `\\12\catcode `\$12\catcode
  `\&12\catcode `\#12\catcode `\^12\catcode `\_12\catcode `\%12\relax}%
\providecommand \@@startlink[1]{}%
\providecommand \@@endlink[0]{}%
\providecommand \url  [0]{\begingroup\@sanitize@url \@url }%
\providecommand \@url [1]{\endgroup\@href {#1}{\urlprefix }}%
\providecommand \urlprefix  [0]{URL }%
\providecommand \Eprint [0]{\href }%
\providecommand \doibase [0]{http://dx.doi.org/}%
\providecommand \selectlanguage [0]{\@gobble}%
\providecommand \bibinfo  [0]{\@secondoftwo}%
\providecommand \bibfield  [0]{\@secondoftwo}%
\providecommand \translation [1]{[#1]}%
\providecommand \BibitemOpen [0]{}%
\providecommand \bibitemStop [0]{}%
\providecommand \bibitemNoStop [0]{.\EOS\space}%
\providecommand \EOS [0]{\spacefactor3000\relax}%
\providecommand \BibitemShut  [1]{\csname bibitem#1\endcsname}%
\let\auto@bib@innerbib\@empty
\bibitem [{\citenamefont {Zanardi}\ and\ \citenamefont
  {Rasetti}(1999)}]{Zanardi199994}%
  \BibitemOpen
  \bibfield  {author} {\bibinfo {author} {\bibfnamefont {P.}~\bibnamefont
  {Zanardi}}\ and\ \bibinfo {author} {\bibfnamefont {M.}~\bibnamefont
  {Rasetti}},\ }\href {\doibase
  http://dx.doi.org/10.1016/S0375-9601(99)00803-8} {\bibfield  {journal}
  {\bibinfo  {journal} {Physics Letters A}\ }\textbf {\bibinfo {volume}
  {264}},\ \bibinfo {pages} {94 } (\bibinfo {year} {1999})}\BibitemShut
  {NoStop}%
\bibitem [{\citenamefont {Pachos}\ and\ \citenamefont
  {Zanardi}(2001)}]{pachos2001quantum}%
  \BibitemOpen
  \bibfield  {author} {\bibinfo {author} {\bibfnamefont {J.}~\bibnamefont
  {Pachos}}\ and\ \bibinfo {author} {\bibfnamefont {P.}~\bibnamefont
  {Zanardi}},\ }\href {https://doi.org/10.1142/s0217979201004836} {\bibfield
  {journal} {\bibinfo  {journal} {International Journal of Modern Physics B}\
  }\textbf {\bibinfo {volume} {15}},\ \bibinfo {pages} {1257} (\bibinfo {year}
  {2001})}\BibitemShut {NoStop}%
\bibitem [{\citenamefont {Ekert}\ \emph {et~al.}(2000)\citenamefont {Ekert},
  \citenamefont {Ericsson}, \citenamefont {Hayden}, \citenamefont {Inamori},
  \citenamefont {Jones}, \citenamefont {Oi},\ and\ \citenamefont
  {Vedral}}]{292c4cfc0e4847d9aeb6240fd0d707b8}%
  \BibitemOpen
  \bibfield  {author} {\bibinfo {author} {\bibfnamefont {A.}~\bibnamefont
  {Ekert}}, \bibinfo {author} {\bibfnamefont {M.}~\bibnamefont {Ericsson}},
  \bibinfo {author} {\bibfnamefont {P.}~\bibnamefont {Hayden}}, \bibinfo
  {author} {\bibfnamefont {H.}~\bibnamefont {Inamori}}, \bibinfo {author}
  {\bibfnamefont {J.}~\bibnamefont {Jones}}, \bibinfo {author} {\bibfnamefont
  {D.}~\bibnamefont {Oi}}, \ and\ \bibinfo {author} {\bibfnamefont
  {V.}~\bibnamefont {Vedral}},\ }\href {\doibase 10.1080/09500340008232177}
  {\bibfield  {journal} {\bibinfo  {journal} {Journal of Modern Optics}\
  }\textbf {\bibinfo {volume} {47}},\ \bibinfo {pages} {2501} (\bibinfo {year}
  {2000})}\BibitemShut {NoStop}%
\bibitem [{\citenamefont {Berry}(1984)}]{Berry08031984}%
  \BibitemOpen
  \bibfield  {author} {\bibinfo {author} {\bibfnamefont {M.~V.}\ \bibnamefont
  {Berry}},\ }\href@noop {} {\bibfield  {journal} {\bibinfo  {journal}
  {Proceedings of the Royal Society of London. A. Mathematical and Physical
  Sciences}\ }\textbf {\bibinfo {volume} {392}},\ \bibinfo {pages} {45}
  (\bibinfo {year} {1984})}\BibitemShut {NoStop}%
\bibitem [{\citenamefont {Wilczek}\ and\ \citenamefont
  {Zee}(1984)}]{PhysRevLett.52.2111}%
  \BibitemOpen
  \bibfield  {author} {\bibinfo {author} {\bibfnamefont {F.}~\bibnamefont
  {Wilczek}}\ and\ \bibinfo {author} {\bibfnamefont {A.}~\bibnamefont {Zee}},\
  }\href {\doibase 10.1103/PhysRevLett.52.2111} {\bibfield  {journal} {\bibinfo
   {journal} {Phys. Rev. Lett.}\ }\textbf {\bibinfo {volume} {52}},\ \bibinfo
  {pages} {2111} (\bibinfo {year} {1984})}\BibitemShut {NoStop}%
\bibitem [{\citenamefont {Aharonov}\ and\ \citenamefont
  {Anandan}(1987)}]{PhysRevLett.58.1593}%
  \BibitemOpen
  \bibfield  {author} {\bibinfo {author} {\bibfnamefont {Y.}~\bibnamefont
  {Aharonov}}\ and\ \bibinfo {author} {\bibfnamefont {J.}~\bibnamefont
  {Anandan}},\ }\href {\doibase 10.1103/PhysRevLett.58.1593} {\bibfield
  {journal} {\bibinfo  {journal} {Phys. Rev. Lett.}\ }\textbf {\bibinfo
  {volume} {58}},\ \bibinfo {pages} {1593} (\bibinfo {year}
  {1987})}\BibitemShut {NoStop}%
\bibitem [{\citenamefont {Anandan}(1988)}]{Anandan1988171}%
  \BibitemOpen
  \bibfield  {author} {\bibinfo {author} {\bibfnamefont {J.}~\bibnamefont
  {Anandan}},\ }\href {\doibase http://dx.doi.org/10.1016/0375-9601(88)91010-9}
  {\bibfield  {journal} {\bibinfo  {journal} {Physics Letters A}\ }\textbf
  {\bibinfo {volume} {133}},\ \bibinfo {pages} {171 } (\bibinfo {year}
  {1988})}\BibitemShut {NoStop}%
\bibitem [{\citenamefont {Samuel}\ and\ \citenamefont
  {Bhandari}(1988)}]{PhysRevLett.60.2339}%
  \BibitemOpen
  \bibfield  {author} {\bibinfo {author} {\bibfnamefont {J.}~\bibnamefont
  {Samuel}}\ and\ \bibinfo {author} {\bibfnamefont {R.}~\bibnamefont
  {Bhandari}},\ }\href {\doibase 10.1103/PhysRevLett.60.2339} {\bibfield
  {journal} {\bibinfo  {journal} {Phys. Rev. Lett.}\ }\textbf {\bibinfo
  {volume} {60}},\ \bibinfo {pages} {2339} (\bibinfo {year}
  {1988})}\BibitemShut {NoStop}%
\bibitem [{\citenamefont {Moore}\ and\ \citenamefont
  {Stedman}(1990)}]{0305-4470-23-11-027}%
  \BibitemOpen
  \bibfield  {author} {\bibinfo {author} {\bibfnamefont {D.~J.}\ \bibnamefont
  {Moore}}\ and\ \bibinfo {author} {\bibfnamefont {G.~E.}\ \bibnamefont
  {Stedman}},\ }\href {http://stacks.iop.org/0305-4470/23/i=11/a=027}
  {\bibfield  {journal} {\bibinfo  {journal} {Journal of Physics A:
  Mathematical and General}\ }\textbf {\bibinfo {volume} {23}},\ \bibinfo
  {pages} {2049} (\bibinfo {year} {1990})}\BibitemShut {NoStop}%
\bibitem [{\citenamefont {Wang}\ \emph {et~al.}(2007)\citenamefont {Wang},
  \citenamefont {Wu}, \citenamefont {Feng}, \citenamefont {Kwek}, \citenamefont
  {Lai}, \citenamefont {Oh},\ and\ \citenamefont
  {Vedral}}]{PhysRevA.76.044303}%
  \BibitemOpen
  \bibfield  {author} {\bibinfo {author} {\bibfnamefont {Z.~S.}\ \bibnamefont
  {Wang}}, \bibinfo {author} {\bibfnamefont {C.}~\bibnamefont {Wu}}, \bibinfo
  {author} {\bibfnamefont {X.-L.}\ \bibnamefont {Feng}}, \bibinfo {author}
  {\bibfnamefont {L.~C.}\ \bibnamefont {Kwek}}, \bibinfo {author}
  {\bibfnamefont {C.~H.}\ \bibnamefont {Lai}}, \bibinfo {author} {\bibfnamefont
  {C.~H.}\ \bibnamefont {Oh}}, \ and\ \bibinfo {author} {\bibfnamefont
  {V.}~\bibnamefont {Vedral}},\ }\href {\doibase 10.1103/PhysRevA.76.044303}
  {\bibfield  {journal} {\bibinfo  {journal} {Phys. Rev. A}\ }\textbf {\bibinfo
  {volume} {76}},\ \bibinfo {pages} {044303} (\bibinfo {year}
  {2007})}\BibitemShut {NoStop}%
\bibitem [{\citenamefont {Xu}\ \emph {et~al.}(2012)\citenamefont {Xu},
  \citenamefont {Zhang}, \citenamefont {Tong}, \citenamefont {Sj\"oqvist},\
  and\ \citenamefont {Kwek}}]{Xu2012}%
  \BibitemOpen
  \bibfield  {author} {\bibinfo {author} {\bibfnamefont {G.~F.}\ \bibnamefont
  {Xu}}, \bibinfo {author} {\bibfnamefont {J.}~\bibnamefont {Zhang}}, \bibinfo
  {author} {\bibfnamefont {D.~M.}\ \bibnamefont {Tong}}, \bibinfo {author}
  {\bibfnamefont {E.}~\bibnamefont {Sj\"oqvist}}, \ and\ \bibinfo {author}
  {\bibfnamefont {L.~C.}\ \bibnamefont {Kwek}},\ }\href {\doibase
  10.1103/PhysRevLett.109.170501} {\bibfield  {journal} {\bibinfo  {journal}
  {Phys. Rev. Lett.}\ }\textbf {\bibinfo {volume} {109}},\ \bibinfo {pages}
  {170501} (\bibinfo {year} {2012})}\BibitemShut {NoStop}%
\bibitem [{\citenamefont {M\"{o}tt\"{o}nen}\ \emph {et~al.}(2008)\citenamefont
  {M\"{o}tt\"{o}nen}, \citenamefont {Vartiainen},\ and\ \citenamefont
  {Pekola}}]{mottonen2008experimental}%
  \BibitemOpen
  \bibfield  {author} {\bibinfo {author} {\bibfnamefont {M.}~\bibnamefont
  {M\"{o}tt\"{o}nen}}, \bibinfo {author} {\bibfnamefont {J.~J.}\ \bibnamefont
  {Vartiainen}}, \ and\ \bibinfo {author} {\bibfnamefont {J.~P.}\ \bibnamefont
  {Pekola}},\ }\href {https://doi.org/10.1103/physrevlett.100.177201}
  {\bibfield  {journal} {\bibinfo  {journal} {Physical Review Letters}\
  }\textbf {\bibinfo {volume} {100}},\ \bibinfo {pages} {177201} (\bibinfo
  {year} {2008})}\BibitemShut {NoStop}%
\bibitem [{\citenamefont {Brosco}\ \emph {et~al.}(2008)\citenamefont {Brosco},
  \citenamefont {Fazio}, \citenamefont {Hekking},\ and\ \citenamefont
  {Joye}}]{brosco2008non}%
  \BibitemOpen
  \bibfield  {author} {\bibinfo {author} {\bibfnamefont {V.}~\bibnamefont
  {Brosco}}, \bibinfo {author} {\bibfnamefont {R.}~\bibnamefont {Fazio}},
  \bibinfo {author} {\bibfnamefont {F.~W.~J.}\ \bibnamefont {Hekking}}, \ and\
  \bibinfo {author} {\bibfnamefont {A.}~\bibnamefont {Joye}},\ }\href
  {https://doi.org/10.1103/physrevlett.100.027002} {\bibfield  {journal}
  {\bibinfo  {journal} {Physical Review Letters}\ }\textbf {\bibinfo {volume}
  {100}},\ \bibinfo {pages} {027002} (\bibinfo {year} {2008})}\BibitemShut
  {NoStop}%
\bibitem [{\citenamefont {Faoro}\ \emph {et~al.}(2003)\citenamefont {Faoro},
  \citenamefont {Siewert},\ and\ \citenamefont {Fazio}}]{faoro2003non}%
  \BibitemOpen
  \bibfield  {author} {\bibinfo {author} {\bibfnamefont {L.}~\bibnamefont
  {Faoro}}, \bibinfo {author} {\bibfnamefont {J.}~\bibnamefont {Siewert}}, \
  and\ \bibinfo {author} {\bibfnamefont {R.}~\bibnamefont {Fazio}},\ }\href
  {https://doi.org/10.1103/physrevlett.90.028301} {\bibfield  {journal}
  {\bibinfo  {journal} {Physical Review Letters}\ }\textbf {\bibinfo {volume}
  {90}},\ \bibinfo {pages} {028301} (\bibinfo {year} {2003})}\BibitemShut
  {NoStop}%
\bibitem [{\citenamefont {Solinas}\ \emph {et~al.}(2010)\citenamefont
  {Solinas}, \citenamefont {Pirkkalainen},\ and\ \citenamefont
  {M\"{o}tt\"{o}nen}}]{solinas2010ground}%
  \BibitemOpen
  \bibfield  {author} {\bibinfo {author} {\bibfnamefont {P.}~\bibnamefont
  {Solinas}}, \bibinfo {author} {\bibfnamefont {J.-M.}\ \bibnamefont
  {Pirkkalainen}}, \ and\ \bibinfo {author} {\bibfnamefont {M.}~\bibnamefont
  {M\"{o}tt\"{o}nen}},\ }\href {https://doi.org/10.1103/physreva.82.052304}
  {\bibfield  {journal} {\bibinfo  {journal} {Physical Review A}\ }\textbf
  {\bibinfo {volume} {82}},\ \bibinfo {pages} {052304} (\bibinfo {year}
  {2010})}\BibitemShut {NoStop}%
\bibitem [{\citenamefont {Pirkkalainen}\ \emph {et~al.}(2010)\citenamefont
  {Pirkkalainen}, \citenamefont {Solinas}, \citenamefont {Pekola},\ and\
  \citenamefont {M\"{o}tt\"{o}nen}}]{pirkkalainen2010non}%
  \BibitemOpen
  \bibfield  {author} {\bibinfo {author} {\bibfnamefont {J.-M.}\ \bibnamefont
  {Pirkkalainen}}, \bibinfo {author} {\bibfnamefont {P.}~\bibnamefont
  {Solinas}}, \bibinfo {author} {\bibfnamefont {J.~P.}\ \bibnamefont {Pekola}},
  \ and\ \bibinfo {author} {\bibfnamefont {M.}~\bibnamefont
  {M\"{o}tt\"{o}nen}},\ }\href {https://doi.org/10.1103/physrevb.81.174506}
  {\bibfield  {journal} {\bibinfo  {journal} {Physical Review B}\ }\textbf
  {\bibinfo {volume} {81}},\ \bibinfo {pages} {174506} (\bibinfo {year}
  {2010})}\BibitemShut {NoStop}%
\bibitem [{\citenamefont {Feng}\ \emph {et~al.}(2013)\citenamefont {Feng},
  \citenamefont {Xu},\ and\ \citenamefont {Long}}]{Feng2013}%
  \BibitemOpen
  \bibfield  {author} {\bibinfo {author} {\bibfnamefont {G.}~\bibnamefont
  {Feng}}, \bibinfo {author} {\bibfnamefont {G.}~\bibnamefont {Xu}}, \ and\
  \bibinfo {author} {\bibfnamefont {G.}~\bibnamefont {Long}},\ }\href {\doibase
  10.1103/PhysRevLett.110.190501} {\bibfield  {journal} {\bibinfo  {journal}
  {Phys. Rev. Lett.}\ }\textbf {\bibinfo {volume} {110}},\ \bibinfo {pages}
  {190501} (\bibinfo {year} {2013})}\BibitemShut {NoStop}%
\bibitem [{\citenamefont {Li}\ \emph {et~al.}(2017)\citenamefont {Li},
  \citenamefont {Liu},\ and\ \citenamefont {Long}}]{Li2017}%
  \BibitemOpen
  \bibfield  {author} {\bibinfo {author} {\bibfnamefont {H.}~\bibnamefont
  {Li}}, \bibinfo {author} {\bibfnamefont {Y.}~\bibnamefont {Liu}}, \ and\
  \bibinfo {author} {\bibfnamefont {G.}~\bibnamefont {Long}},\ }\href {\doibase
  10.1007/s11433-017-9058-7} {\bibfield  {journal} {\bibinfo  {journal}
  {Science China Physics, Mechanics {\&} Astronomy}\ }\textbf {\bibinfo
  {volume} {60}},\ \bibinfo {pages} {080311} (\bibinfo {year}
  {2017})}\BibitemShut {NoStop}%
\bibitem [{\citenamefont {Pechal}\ \emph {et~al.}(2012)\citenamefont {Pechal},
  \citenamefont {Berger}, \citenamefont {Abdumalikov}, \citenamefont {Fink},
  \citenamefont {Mlynek}, \citenamefont {Steffen}, \citenamefont {Wallraff},\
  and\ \citenamefont {Filipp}}]{pechal2012geometric}%
  \BibitemOpen
  \bibfield  {author} {\bibinfo {author} {\bibfnamefont {M.}~\bibnamefont
  {Pechal}}, \bibinfo {author} {\bibfnamefont {S.}~\bibnamefont {Berger}},
  \bibinfo {author} {\bibfnamefont {A.~A.}\ \bibnamefont {Abdumalikov}},
  \bibinfo {author} {\bibfnamefont {J.~M.}\ \bibnamefont {Fink}}, \bibinfo
  {author} {\bibfnamefont {J.~A.}\ \bibnamefont {Mlynek}}, \bibinfo {author}
  {\bibfnamefont {L.}~\bibnamefont {Steffen}}, \bibinfo {author} {\bibfnamefont
  {A.}~\bibnamefont {Wallraff}}, \ and\ \bibinfo {author} {\bibfnamefont
  {S.}~\bibnamefont {Filipp}},\ }\href
  {https://doi.org/10.1103/physrevlett.108.170401} {\bibfield  {journal}
  {\bibinfo  {journal} {Physical Review Letters}\ }\textbf {\bibinfo {volume}
  {108}},\ \bibinfo {pages} {170401} (\bibinfo {year} {2012})}\BibitemShut
  {NoStop}%
\bibitem [{\citenamefont {Duan}\ \emph {et~al.}(2001)\citenamefont {Duan},
  \citenamefont {Cirac},\ and\ \citenamefont {Zoller}}]{duan2001geometric}%
  \BibitemOpen
  \bibfield  {author} {\bibinfo {author} {\bibfnamefont {L.-M.}\ \bibnamefont
  {Duan}}, \bibinfo {author} {\bibfnamefont {J.~I.}\ \bibnamefont {Cirac}}, \
  and\ \bibinfo {author} {\bibfnamefont {P.}~\bibnamefont {Zoller}},\ }\href
  {https://doi.org/10.1126/science.1058835} {\bibfield  {journal} {\bibinfo
  {journal} {Science}\ }\textbf {\bibinfo {volume} {292}},\ \bibinfo {pages}
  {1695} (\bibinfo {year} {2001})}\BibitemShut {NoStop}%
\bibitem [{\citenamefont {Unanyan}\ \emph {et~al.}(1999)\citenamefont
  {Unanyan}, \citenamefont {Shore},\ and\ \citenamefont
  {Bergmann}}]{unanyan1999laser}%
  \BibitemOpen
  \bibfield  {author} {\bibinfo {author} {\bibfnamefont {R.~G.}\ \bibnamefont
  {Unanyan}}, \bibinfo {author} {\bibfnamefont {B.~W.}\ \bibnamefont {Shore}},
  \ and\ \bibinfo {author} {\bibfnamefont {K.}~\bibnamefont {Bergmann}},\
  }\href {https://doi.org/10.1103/physreva.59.2910} {\bibfield  {journal}
  {\bibinfo  {journal} {Physical Review A}\ }\textbf {\bibinfo {volume} {59}},\
  \bibinfo {pages} {2910} (\bibinfo {year} {1999})}\BibitemShut {NoStop}%
\bibitem [{\citenamefont {{Abdumalikov Jr}}\ \emph {et~al.}(2013)\citenamefont
  {{Abdumalikov Jr}}, \citenamefont {Fink}, \citenamefont {Juliusson},
  \citenamefont {Pechal}, \citenamefont {Berger}, \citenamefont {Wallraff},\
  and\ \citenamefont {Filipp}}]{AbdumalikovJr2013}%
  \BibitemOpen
  \bibfield  {author} {\bibinfo {author} {\bibfnamefont {A.~A.}\ \bibnamefont
  {{Abdumalikov Jr}}}, \bibinfo {author} {\bibfnamefont {J.~M.}\ \bibnamefont
  {Fink}}, \bibinfo {author} {\bibfnamefont {K.}~\bibnamefont {Juliusson}},
  \bibinfo {author} {\bibfnamefont {M.}~\bibnamefont {Pechal}}, \bibinfo
  {author} {\bibfnamefont {S.}~\bibnamefont {Berger}}, \bibinfo {author}
  {\bibfnamefont {A.}~\bibnamefont {Wallraff}}, \ and\ \bibinfo {author}
  {\bibfnamefont {S.}~\bibnamefont {Filipp}},\ }\href {\doibase
  10.1038/nature12010} {\bibfield  {journal} {\bibinfo  {journal} {Nature}\
  }\textbf {\bibinfo {volume} {496}},\ \bibinfo {pages} {482} (\bibinfo {year}
  {2013})}\BibitemShut {NoStop}%
\bibitem [{\citenamefont {Arroyo-Camejo}\ \emph {et~al.}(2014)\citenamefont
  {Arroyo-Camejo}, \citenamefont {Lazariev}, \citenamefont {Hell},\ and\
  \citenamefont {Balasubramanian}}]{ArroyoCamejo2014}%
  \BibitemOpen
  \bibfield  {author} {\bibinfo {author} {\bibfnamefont {S.}~\bibnamefont
  {Arroyo-Camejo}}, \bibinfo {author} {\bibfnamefont {A.}~\bibnamefont
  {Lazariev}}, \bibinfo {author} {\bibfnamefont {S.~W.}\ \bibnamefont {Hell}},
  \ and\ \bibinfo {author} {\bibfnamefont {G.}~\bibnamefont
  {Balasubramanian}},\ }\href {https://doi.org/10.1038/ncomms5870} {\bibfield
  {journal} {\bibinfo  {journal} {Nature Communications}\ }\textbf {\bibinfo
  {volume} {5}},\ \bibinfo {pages} {4870} (\bibinfo {year} {2014})}\BibitemShut
  {NoStop}%
\bibitem [{\citenamefont {Zu}\ \emph {et~al.}(2014)\citenamefont {Zu},
  \citenamefont {Wang}, \citenamefont {He}, \citenamefont {Zhang},
  \citenamefont {Dai}, \citenamefont {Wang},\ and\ \citenamefont
  {Duan}}]{Zu2014}%
  \BibitemOpen
  \bibfield  {author} {\bibinfo {author} {\bibfnamefont {C.}~\bibnamefont
  {Zu}}, \bibinfo {author} {\bibfnamefont {W.-B.}\ \bibnamefont {Wang}},
  \bibinfo {author} {\bibfnamefont {L.}~\bibnamefont {He}}, \bibinfo {author}
  {\bibfnamefont {W.-G.}\ \bibnamefont {Zhang}}, \bibinfo {author}
  {\bibfnamefont {C.-Y.}\ \bibnamefont {Dai}}, \bibinfo {author} {\bibfnamefont
  {F.}~\bibnamefont {Wang}}, \ and\ \bibinfo {author} {\bibfnamefont {L.-M.}\
  \bibnamefont {Duan}},\ }\href {\doibase 10.1038/nature13729} {\bibfield
  {journal} {\bibinfo  {journal} {Nature}\ }\textbf {\bibinfo {volume} {514}},\
  \bibinfo {pages} {72} (\bibinfo {year} {2014})}\BibitemShut {NoStop}%
\bibitem [{\citenamefont {Golovach}\ \emph {et~al.}(2010)\citenamefont
  {Golovach}, \citenamefont {Borhani},\ and\ \citenamefont
  {Loss}}]{golovach2010holonomic}%
  \BibitemOpen
  \bibfield  {author} {\bibinfo {author} {\bibfnamefont {V.~N.}\ \bibnamefont
  {Golovach}}, \bibinfo {author} {\bibfnamefont {M.}~\bibnamefont {Borhani}}, \
  and\ \bibinfo {author} {\bibfnamefont {D.}~\bibnamefont {Loss}},\ }\href
  {https://doi.org/10.1103/physreva.81.022315} {\bibfield  {journal} {\bibinfo
  {journal} {Physical Review A}\ }\textbf {\bibinfo {volume} {81}},\ \bibinfo
  {pages} {022315} (\bibinfo {year} {2010})}\BibitemShut {NoStop}%
\bibitem [{\citenamefont {Sekiguchi}\ \emph {et~al.}(2017)\citenamefont
  {Sekiguchi}, \citenamefont {Niikura}, \citenamefont {Kuroiwa}, \citenamefont
  {Kano},\ and\ \citenamefont {Kosaka}}]{Sekiguchi2017}%
  \BibitemOpen
  \bibfield  {author} {\bibinfo {author} {\bibfnamefont {Y.}~\bibnamefont
  {Sekiguchi}}, \bibinfo {author} {\bibfnamefont {N.}~\bibnamefont {Niikura}},
  \bibinfo {author} {\bibfnamefont {R.}~\bibnamefont {Kuroiwa}}, \bibinfo
  {author} {\bibfnamefont {H.}~\bibnamefont {Kano}}, \ and\ \bibinfo {author}
  {\bibfnamefont {H.}~\bibnamefont {Kosaka}},\ }\href {\doibase
  10.1038/nphoton.2017.40} {\bibfield  {journal} {\bibinfo  {journal} {Nature
  Photonics}\ }\textbf {\bibinfo {volume} {11}},\ \bibinfo {pages} {309}
  (\bibinfo {year} {2017})}\BibitemShut {NoStop}%
\bibitem [{\citenamefont {Carollo}\ \emph {et~al.}(2003)\citenamefont
  {Carollo}, \citenamefont {Fuentes-Guridi}, \citenamefont {Santos},\ and\
  \citenamefont {Vedral}}]{PhysRevLett.90.160402}%
  \BibitemOpen
  \bibfield  {author} {\bibinfo {author} {\bibfnamefont {A.}~\bibnamefont
  {Carollo}}, \bibinfo {author} {\bibfnamefont {I.}~\bibnamefont
  {Fuentes-Guridi}}, \bibinfo {author} {\bibfnamefont {M.~F.}\ \bibnamefont
  {Santos}}, \ and\ \bibinfo {author} {\bibfnamefont {V.}~\bibnamefont
  {Vedral}},\ }\href {\doibase 10.1103/PhysRevLett.90.160402} {\bibfield
  {journal} {\bibinfo  {journal} {Phys. Rev. Lett.}\ }\textbf {\bibinfo
  {volume} {90}},\ \bibinfo {pages} {160402} (\bibinfo {year}
  {2003})}\BibitemShut {NoStop}%
\bibitem [{\citenamefont {Whitney}\ and\ \citenamefont
  {Gefen}(2003)}]{PhysRevLett.90.190402}%
  \BibitemOpen
  \bibfield  {author} {\bibinfo {author} {\bibfnamefont {R.~S.}\ \bibnamefont
  {Whitney}}\ and\ \bibinfo {author} {\bibfnamefont {Y.}~\bibnamefont
  {Gefen}},\ }\href {\doibase 10.1103/PhysRevLett.90.190402} {\bibfield
  {journal} {\bibinfo  {journal} {Phys. Rev. Lett.}\ }\textbf {\bibinfo
  {volume} {90}},\ \bibinfo {pages} {190402} (\bibinfo {year}
  {2003})}\BibitemShut {NoStop}%
\bibitem [{\citenamefont {Li}\ and\ \citenamefont
  {Li}(2011)}]{1751-8121-44-9-095304}%
  \BibitemOpen
  \bibfield  {author} {\bibinfo {author} {\bibfnamefont {X.}~\bibnamefont
  {Li}}\ and\ \bibinfo {author} {\bibfnamefont {Z.}~\bibnamefont {Li}},\ }\href
  {http://stacks.iop.org/1751-8121/44/i=9/a=095304} {\bibfield  {journal}
  {\bibinfo  {journal} {Journal of Physics A: Mathematical and Theoretical}\
  }\textbf {\bibinfo {volume} {44}},\ \bibinfo {pages} {095304} (\bibinfo
  {year} {2011})}\BibitemShut {NoStop}%
\bibitem [{\citenamefont {Fuentes-Guridi}\ \emph {et~al.}(2005)\citenamefont
  {Fuentes-Guridi}, \citenamefont {Girelli},\ and\ \citenamefont
  {Livine}}]{FuentesGuridi2005}%
  \BibitemOpen
  \bibfield  {author} {\bibinfo {author} {\bibfnamefont {I.}~\bibnamefont
  {Fuentes-Guridi}}, \bibinfo {author} {\bibfnamefont {F.}~\bibnamefont
  {Girelli}}, \ and\ \bibinfo {author} {\bibfnamefont {E.}~\bibnamefont
  {Livine}},\ }\href {\doibase 10.1103/PhysRevLett.94.020503} {\bibfield
  {journal} {\bibinfo  {journal} {Phys. Rev. Lett.}\ }\textbf {\bibinfo
  {volume} {94}},\ \bibinfo {pages} {45} (\bibinfo {year} {2005})}\BibitemShut
  {NoStop}%
\bibitem [{\citenamefont {Thunstr\"om}\ \emph {et~al.}(2005)\citenamefont
  {Thunstr\"om}, \citenamefont {{\AA}berg},\ and\ \citenamefont
  {Sj\"oqvist}}]{Thunstrom2005}%
  \BibitemOpen
  \bibfield  {author} {\bibinfo {author} {\bibfnamefont {P.}~\bibnamefont
  {Thunstr\"om}}, \bibinfo {author} {\bibfnamefont {J.}~\bibnamefont
  {{\AA}berg}}, \ and\ \bibinfo {author} {\bibfnamefont {E.}~\bibnamefont
  {Sj\"oqvist}},\ }\href {http://doi.org/10.1103/PhysRevA.72.022328} {\bibfield
   {journal} {\bibinfo  {journal} {Phys. Rev. A}\ }\textbf {\bibinfo {volume}
  {72}},\ \bibinfo {pages} {022328} (\bibinfo {year} {2005})}\BibitemShut
  {NoStop}%
\bibitem [{\citenamefont {Sarandy}\ and\ \citenamefont
  {Lidar}(2006)}]{Sarandy2006}%
  \BibitemOpen
  \bibfield  {author} {\bibinfo {author} {\bibfnamefont {M.~S.}\ \bibnamefont
  {Sarandy}}\ and\ \bibinfo {author} {\bibfnamefont {D.~A.}\ \bibnamefont
  {Lidar}},\ }\href {\doibase 10.1103/PhysRevA.73.062101} {\bibfield  {journal}
  {\bibinfo  {journal} {Phys. Rev. A}\ }\textbf {\bibinfo {volume} {73}},\
  \bibinfo {pages} {557} (\bibinfo {year} {2006})}\BibitemShut {NoStop}%
\bibitem [{\citenamefont {M{\o}ller}\ \emph {et~al.}(2008)\citenamefont
  {M{\o}ller}, \citenamefont {Madsen},\ and\ \citenamefont
  {M{\o}lmer}}]{Moller2008}%
  \BibitemOpen
  \bibfield  {author} {\bibinfo {author} {\bibfnamefont {D.}~\bibnamefont
  {M{\o}ller}}, \bibinfo {author} {\bibfnamefont {L.~B.}\ \bibnamefont
  {Madsen}}, \ and\ \bibinfo {author} {\bibfnamefont {K.}~\bibnamefont
  {M{\o}lmer}},\ }\href {http://doi.org/10.1103/PhysRevA.77.022306} {\bibfield
  {journal} {\bibinfo  {journal} {Phys. Rev. A}\ }\textbf {\bibinfo {volume}
  {77}},\ \bibinfo {pages} {022306} (\bibinfo {year} {2008})}\BibitemShut
  {NoStop}%
\bibitem [{\citenamefont {Shenvi}\ \emph {et~al.}(2003)\citenamefont {Shenvi},
  \citenamefont {Brown},\ and\ \citenamefont {Whaley}}]{She.Bro.Wha:03}%
  \BibitemOpen
  \bibfield  {author} {\bibinfo {author} {\bibfnamefont {N.}~\bibnamefont
  {Shenvi}}, \bibinfo {author} {\bibfnamefont {K.~R.}\ \bibnamefont {Brown}}, \
  and\ \bibinfo {author} {\bibfnamefont {K.~B.}\ \bibnamefont {Whaley}},\
  }\href {\doibase 10.1103/PhysRevA.68.052313} {\bibfield  {journal} {\bibinfo
  {journal} {Phys. Rev. A}\ }\textbf {\bibinfo {volume} {68}},\ \bibinfo
  {pages} {052313} (\bibinfo {year} {2003})}\BibitemShut {NoStop}%
\bibitem [{\citenamefont {De~Chiara}\ and\ \citenamefont
  {Palma}(2008)}]{deChiara:2008}%
  \BibitemOpen
  \bibfield  {author} {\bibinfo {author} {\bibfnamefont {G.}~\bibnamefont
  {De~Chiara}}\ and\ \bibinfo {author} {\bibfnamefont {G.~M.}\ \bibnamefont
  {Palma}},\ }\href {\doibase 10.1007/s10773-007-9575-z} {\bibfield  {journal}
  {\bibinfo  {journal} {International Journal of Theoretical Physics}\ }\textbf
  {\bibinfo {volume} {47}},\ \bibinfo {pages} {2165} (\bibinfo {year}
  {2008})}\BibitemShut {NoStop}%
\bibitem [{\citenamefont {De~Chiara}\ and\ \citenamefont
  {Palma}(2003)}]{PhysRevLett.91.090404}%
  \BibitemOpen
  \bibfield  {author} {\bibinfo {author} {\bibfnamefont {G.}~\bibnamefont
  {De~Chiara}}\ and\ \bibinfo {author} {\bibfnamefont {G.~M.}\ \bibnamefont
  {Palma}},\ }\href {\doibase 10.1103/PhysRevLett.91.090404} {\bibfield
  {journal} {\bibinfo  {journal} {Phys. Rev. Lett.}\ }\textbf {\bibinfo
  {volume} {91}},\ \bibinfo {pages} {090404} (\bibinfo {year}
  {2003})}\BibitemShut {NoStop}%
\bibitem [{\citenamefont {Solinas}\ \emph {et~al.}(2004)\citenamefont
  {Solinas}, \citenamefont {Zanardi},\ and\ \citenamefont
  {Zangh{\`{\i}}}}]{solinas2004robustness}%
  \BibitemOpen
  \bibfield  {author} {\bibinfo {author} {\bibfnamefont {P.}~\bibnamefont
  {Solinas}}, \bibinfo {author} {\bibfnamefont {P.}~\bibnamefont {Zanardi}}, \
  and\ \bibinfo {author} {\bibfnamefont {N.}~\bibnamefont {Zangh{\`{\i}}}},\
  }\href {https://doi.org/10.1103/physreva.70.042316} {\bibfield  {journal}
  {\bibinfo  {journal} {Physical Review A}\ }\textbf {\bibinfo {volume} {70}},\
  \bibinfo {pages} {042316} (\bibinfo {year} {2004})}\BibitemShut {NoStop}%
\bibitem [{\citenamefont {Solinas}\ \emph {et~al.}(2012)\citenamefont
  {Solinas}, \citenamefont {Sassetti}, \citenamefont {Truini},\ and\
  \citenamefont {Zangh{\`{\i}}}}]{solinas2012stability}%
  \BibitemOpen
  \bibfield  {author} {\bibinfo {author} {\bibfnamefont {P.}~\bibnamefont
  {Solinas}}, \bibinfo {author} {\bibfnamefont {M.}~\bibnamefont {Sassetti}},
  \bibinfo {author} {\bibfnamefont {P.}~\bibnamefont {Truini}}, \ and\ \bibinfo
  {author} {\bibfnamefont {N.}~\bibnamefont {Zangh{\`{\i}}}},\ }\href
  {https://doi.org/10.1088/1367-2630/14/9/093006} {\bibfield  {journal}
  {\bibinfo  {journal} {New Journal of Physics}\ }\textbf {\bibinfo {volume}
  {14}},\ \bibinfo {pages} {093006} (\bibinfo {year} {2012})}\BibitemShut
  {NoStop}%
\bibitem [{\citenamefont {Shor}(1994)}]{Sho:94}%
  \BibitemOpen
  \bibfield  {author} {\bibinfo {author} {\bibfnamefont {P.}~\bibnamefont
  {Shor}},\ }in\ \href {https://doi.org/10.1109/sfcs.1994.365700} {\emph
  {\bibinfo {booktitle} {Proceedings 35th Annual Symposium on Foundations of
  Computer Science}}}\ (\bibinfo  {publisher} {{IEEE} Comput. Soc. Press},\
  \bibinfo {year} {1994})\ pp.\ \bibinfo {pages} {124--134}\BibitemShut
  {NoStop}%
\bibitem [{\citenamefont {Grover}(1996)}]{Gro:96}%
  \BibitemOpen
  \bibfield  {author} {\bibinfo {author} {\bibfnamefont {L.~K.}\ \bibnamefont
  {Grover}},\ }in\ \href {https://doi.org/10.1145/237814.237866} {\emph
  {\bibinfo {booktitle} {Proceedings of the twenty-eighth annual {ACM}
  symposium on Theory of computing - {STOC} 96}}}\ (\bibinfo  {publisher}
  {{ACM} Press},\ \bibinfo {year} {1996})\ pp.\ \bibinfo {pages}
  {212--219}\BibitemShut {NoStop}%
\bibitem [{\citenamefont {Long}\ \emph {et~al.}(2000)\citenamefont {Long},
  \citenamefont {Li}, \citenamefont {Zhang},\ and\ \citenamefont
  {Tu}}]{Lon.Li.Zha.Tu:00}%
  \BibitemOpen
  \bibfield  {author} {\bibinfo {author} {\bibfnamefont {G.~L.}\ \bibnamefont
  {Long}}, \bibinfo {author} {\bibfnamefont {Y.~S.}\ \bibnamefont {Li}},
  \bibinfo {author} {\bibfnamefont {W.~L.}\ \bibnamefont {Zhang}}, \ and\
  \bibinfo {author} {\bibfnamefont {C.~C.}\ \bibnamefont {Tu}},\ }\href
  {\doibase 10.1103/PhysRevA.61.042305} {\bibfield  {journal} {\bibinfo
  {journal} {Phys. Rev. A}\ }\textbf {\bibinfo {volume} {61}},\ \bibinfo
  {pages} {042305} (\bibinfo {year} {2000})}\BibitemShut {NoStop}%
\bibitem [{\citenamefont {Guo}\ \emph {et~al.}(2001)\citenamefont {Guo},
  \citenamefont {Long},\ and\ \citenamefont {Sun}}]{Guo2001}%
  \BibitemOpen
  \bibfield  {author} {\bibinfo {author} {\bibfnamefont {H.}~\bibnamefont
  {Guo}}, \bibinfo {author} {\bibfnamefont {G.-L.}\ \bibnamefont {Long}}, \
  and\ \bibinfo {author} {\bibfnamefont {Y.}~\bibnamefont {Sun}},\ }\href
  {\doibase 10.1002/jccs.200100067} {\bibfield  {journal} {\bibinfo  {journal}
  {Journal of the Chinese Chemical Society}\ }\textbf {\bibinfo {volume}
  {48}},\ \bibinfo {pages} {449} (\bibinfo {year} {2001})}\BibitemShut
  {NoStop}%
\bibitem [{\citenamefont {Niwa}\ \emph {et~al.}(2002)\citenamefont {Niwa},
  \citenamefont {Matsumoto},\ and\ \citenamefont {Imai}}]{Niwa2002}%
  \BibitemOpen
  \bibfield  {author} {\bibinfo {author} {\bibfnamefont {J.}~\bibnamefont
  {Niwa}}, \bibinfo {author} {\bibfnamefont {K.}~\bibnamefont {Matsumoto}}, \
  and\ \bibinfo {author} {\bibfnamefont {H.}~\bibnamefont {Imai}},\ }\href
  {\doibase 10.1103/PhysRevA.66.062317} {\bibfield  {journal} {\bibinfo
  {journal} {Phys. Rev. A}\ }\textbf {\bibinfo {volume} {66}},\ \bibinfo
  {pages} {062317} (\bibinfo {year} {2002})}\BibitemShut {NoStop}%
\bibitem [{\citenamefont {Salas}(2008)}]{Salas2007}%
  \BibitemOpen
  \bibfield  {author} {\bibinfo {author} {\bibfnamefont {P.~J.}\ \bibnamefont
  {Salas}},\ }\href {\doibase 10.1140/epjd/e2007-00295-1} {\bibfield  {journal}
  {\bibinfo  {journal} {The European Physical Journal D}\ }\textbf {\bibinfo
  {volume} {46}},\ \bibinfo {pages} {365} (\bibinfo {year} {2008})}\BibitemShut
  {NoStop}%
\bibitem [{\citenamefont {Filipp}(2008)}]{filipp:2008}%
  \BibitemOpen
  \bibfield  {author} {\bibinfo {author} {\bibfnamefont {S.}~\bibnamefont
  {Filipp}},\ }\href {\doibase 10.1140/epjst/e2008-00720-1} {\bibfield
  {journal} {\bibinfo  {journal} {The European Physical Journal Special
  Topics}\ }\textbf {\bibinfo {volume} {160}},\ \bibinfo {pages} {165}
  (\bibinfo {year} {2008})}\BibitemShut {NoStop}%
\bibitem [{\citenamefont {Filipp}\ \emph {et~al.}(2009)\citenamefont {Filipp},
  \citenamefont {Klepp}, \citenamefont {Hasegawa}, \citenamefont
  {Plonka-Spehr}, \citenamefont {Schmidt}, \citenamefont {Geltenbort},\ and\
  \citenamefont {Rauch}}]{PhysRevLett.102.030404}%
  \BibitemOpen
  \bibfield  {author} {\bibinfo {author} {\bibfnamefont {S.}~\bibnamefont
  {Filipp}}, \bibinfo {author} {\bibfnamefont {J.}~\bibnamefont {Klepp}},
  \bibinfo {author} {\bibfnamefont {Y.}~\bibnamefont {Hasegawa}}, \bibinfo
  {author} {\bibfnamefont {C.}~\bibnamefont {Plonka-Spehr}}, \bibinfo {author}
  {\bibfnamefont {U.}~\bibnamefont {Schmidt}}, \bibinfo {author} {\bibfnamefont
  {P.}~\bibnamefont {Geltenbort}}, \ and\ \bibinfo {author} {\bibfnamefont
  {H.}~\bibnamefont {Rauch}},\ }\href {\doibase 10.1103/PhysRevLett.102.030404}
  {\bibfield  {journal} {\bibinfo  {journal} {Phys. Rev. Lett.}\ }\textbf
  {\bibinfo {volume} {102}},\ \bibinfo {pages} {030404} (\bibinfo {year}
  {2009})}\BibitemShut {NoStop}%
\bibitem [{\citenamefont {Berger}\ \emph {et~al.}(2013)\citenamefont {Berger},
  \citenamefont {Pechal}, \citenamefont {Abdumalikov}, \citenamefont {Eichler},
  \citenamefont {Steffen}, \citenamefont {Fedorov}, \citenamefont {Wallraff},\
  and\ \citenamefont {Filipp}}]{PhysRevA.87.060303}%
  \BibitemOpen
  \bibfield  {author} {\bibinfo {author} {\bibfnamefont {S.}~\bibnamefont
  {Berger}}, \bibinfo {author} {\bibfnamefont {M.}~\bibnamefont {Pechal}},
  \bibinfo {author} {\bibfnamefont {A.~A.}\ \bibnamefont {Abdumalikov}},
  \bibinfo {author} {\bibfnamefont {C.}~\bibnamefont {Eichler}}, \bibinfo
  {author} {\bibfnamefont {L.}~\bibnamefont {Steffen}}, \bibinfo {author}
  {\bibfnamefont {A.}~\bibnamefont {Fedorov}}, \bibinfo {author} {\bibfnamefont
  {A.}~\bibnamefont {Wallraff}}, \ and\ \bibinfo {author} {\bibfnamefont
  {S.}~\bibnamefont {Filipp}},\ }\href {\doibase 10.1103/PhysRevA.87.060303}
  {\bibfield  {journal} {\bibinfo  {journal} {Phys. Rev. A}\ }\textbf {\bibinfo
  {volume} {87}},\ \bibinfo {pages} {060303} (\bibinfo {year}
  {2013})}\BibitemShut {NoStop}%
\bibitem [{\citenamefont {Aguilar}\ \emph {et~al.}(2016)\citenamefont
  {Aguilar}, \citenamefont {Chryssomalakos},\ and\ \citenamefont
  {Guzm{\'{a}}n}}]{Agu.Chr.Guz:15a}%
  \BibitemOpen
  \bibfield  {author} {\bibinfo {author} {\bibfnamefont {P.}~\bibnamefont
  {Aguilar}}, \bibinfo {author} {\bibfnamefont {C.}~\bibnamefont
  {Chryssomalakos}}, \ and\ \bibinfo {author} {\bibfnamefont {E.}~\bibnamefont
  {Guzm{\'{a}}n}},\ }\href {https://doi.org/10.1142/s021773231650098x}
  {\bibfield  {journal} {\bibinfo  {journal} {Modern Physics Letters A}\
  }\textbf {\bibinfo {volume} {31}},\ \bibinfo {pages} {1650098} (\bibinfo
  {year} {2016})}\BibitemShut {NoStop}%
\bibitem [{\citenamefont {Zee}(1988)}]{PhysRevA.38.1}%
  \BibitemOpen
  \bibfield  {author} {\bibinfo {author} {\bibfnamefont {A.}~\bibnamefont
  {Zee}},\ }\href {http://link.aps.org/doi/10.1103/PhysRevA.38.1} {\bibfield
  {journal} {\bibinfo  {journal} {Phys. Rev. A}\ }\textbf {\bibinfo {volume}
  {38}},\ \bibinfo {pages} {1} (\bibinfo {year} {1988})}\BibitemShut {NoStop}%
\bibitem [{\citenamefont {Tycko}(1987)}]{Tycko1987}%
  \BibitemOpen
  \bibfield  {author} {\bibinfo {author} {\bibfnamefont {R.}~\bibnamefont
  {Tycko}},\ }\href {\doibase 10.1103/PhysRevLett.58.2281} {\bibfield
  {journal} {\bibinfo  {journal} {Phys. Rev. Lett.}\ }\textbf {\bibinfo
  {volume} {58}},\ \bibinfo {pages} {2281} (\bibinfo {year}
  {1987})}\BibitemShut {NoStop}%
\bibitem [{\citenamefont {Messiah}(1962)}]{messiah1962}%
  \BibitemOpen
  \bibfield  {author} {\bibinfo {author} {\bibfnamefont {A.}~\bibnamefont
  {Messiah}},\ }\href@noop {} {\emph {\bibinfo {title} {Quantum Mechanics}}},\
  Vol.~\bibinfo {volume} {2}\ (\bibinfo  {publisher} {North Holland Publishing
  Company, Amsterdam},\ \bibinfo {year} {1962})\BibitemShut {NoStop}%
\bibitem [{\citenamefont {Peres}(1995)}]{peres:1995}%
  \BibitemOpen
  \bibfield  {author} {\bibinfo {author} {\bibfnamefont {A.}~\bibnamefont
  {Peres}},\ }\href@noop {} {\emph {\bibinfo {title} {Quantum Theory: Concepts
  and Methods}}},\ Fundamental Theories of Physics\ (\bibinfo  {publisher}
  {Kluwer Academic Publishers},\ \bibinfo {year} {1995})\BibitemShut {NoStop}%
\bibitem [{\citenamefont {Bohm}\ \emph {et~al.}(2003)\citenamefont {Bohm},
  \citenamefont {Mostafazadeh}, \citenamefont {Koizumi}, \citenamefont {Niu},\
  and\ \citenamefont {Zwanziger}}]{bohm:2003}%
  \BibitemOpen
  \bibfield  {author} {\bibinfo {author} {\bibfnamefont {A.}~\bibnamefont
  {Bohm}}, \bibinfo {author} {\bibfnamefont {A.}~\bibnamefont {Mostafazadeh}},
  \bibinfo {author} {\bibfnamefont {H.}~\bibnamefont {Koizumi}}, \bibinfo
  {author} {\bibfnamefont {Q.}~\bibnamefont {Niu}}, \ and\ \bibinfo {author}
  {\bibfnamefont {J.}~\bibnamefont {Zwanziger}},\ }\href@noop {} {\emph
  {\bibinfo {title} {The Geometric Phase in Quantum Systems: Foundations,
  Mathematical Concepts, and Applications in Molecular and Condensed Matter
  Physics.}}}\ (\bibinfo  {publisher} {Springer-Verlag},\ \bibinfo {year}
  {2003})\BibitemShut {NoStop}%
\bibitem [{\citenamefont {Chru\'sci\'nski}\ and\ \citenamefont
  {Jamio\l{}kowski}(2004)}]{dariusz:2004}%
  \BibitemOpen
  \bibfield  {author} {\bibinfo {author} {\bibfnamefont {D.}~\bibnamefont
  {Chru\'sci\'nski}}\ and\ \bibinfo {author} {\bibfnamefont {A.}~\bibnamefont
  {Jamio\l{}kowski}},\ }\href@noop {} {\emph {\bibinfo {title} {Geometric
  Phases in Classical and Quantum Mechanics.}}}\ (\bibinfo  {publisher}
  {Birkh\"uaser},\ \bibinfo {year} {2004})\BibitemShut {NoStop}%
\bibitem [{\citenamefont {Jacobs}(2010)}]{jacobs:2010}%
  \BibitemOpen
  \bibfield  {author} {\bibinfo {author} {\bibfnamefont {K.}~\bibnamefont
  {Jacobs}},\ }\href@noop {} {\emph {\bibinfo {title} {Stochastic Processes for
  Physicists Understanding Noisy Systems}}}\ (\bibinfo  {publisher} {Cambridge
  University Press},\ \bibinfo {year} {2010})\BibitemShut {NoStop}%
\bibitem [{\citenamefont {Nakahara}(1990)}]{Nak:90}%
  \BibitemOpen
  \bibfield  {author} {\bibinfo {author} {\bibfnamefont {M.}~\bibnamefont
  {Nakahara}},\ }\href@noop {} {\emph {\bibinfo {title} {Geometry, topology and
  physics}}}\ (\bibinfo  {publisher} {Institute of Physics Publishing},\
  \bibinfo {year} {1990})\BibitemShut {NoStop}%
\bibitem [{\citenamefont {Bengtsson}\ and\ \citenamefont
  {Zyczkowski}(2008)}]{Bengtsson2008}%
  \BibitemOpen
  \bibfield  {author} {\bibinfo {author} {\bibfnamefont {I.}~\bibnamefont
  {Bengtsson}}\ and\ \bibinfo {author} {\bibfnamefont {K.}~\bibnamefont
  {Zyczkowski}},\ }\href@noop {} {\emph {\bibinfo {title} {Geometry of Quantum
  States: An Introduction to Quantum Entanglement}}}\ (\bibinfo  {publisher}
  {Cambridge University Press},\ \bibinfo {year} {2008})\BibitemShut {NoStop}%
\bibitem [{\citenamefont {Fuentes-Guridi}\ \emph {et~al.}(2002)\citenamefont
  {Fuentes-Guridi}, \citenamefont {Carollo}, \citenamefont {Bose},\ and\
  \citenamefont {Vedral}}]{Fue.Car.Bos.Ved:02}%
  \BibitemOpen
  \bibfield  {author} {\bibinfo {author} {\bibfnamefont {I.}~\bibnamefont
  {Fuentes-Guridi}}, \bibinfo {author} {\bibfnamefont {A.}~\bibnamefont
  {Carollo}}, \bibinfo {author} {\bibfnamefont {S.}~\bibnamefont {Bose}}, \
  and\ \bibinfo {author} {\bibfnamefont {V.}~\bibnamefont {Vedral}},\ }\href
  {\doibase 10.1103/physrevlett.89.220404} {\bibfield  {journal} {\bibinfo
  {journal} {Phys.{} Rev.{} Lett.{}}\ }\textbf {\bibinfo {volume} {89}},\
  \bibinfo {pages} {220404} (\bibinfo {year} {2002})}\BibitemShut {NoStop}%
\bibitem [{\citenamefont {Liu}\ \emph {et~al.}(2011)\citenamefont {Liu},
  \citenamefont {Feng},\ and\ \citenamefont {Wang}}]{Liu.Fen.Wan:11}%
  \BibitemOpen
  \bibfield  {author} {\bibinfo {author} {\bibfnamefont {T.}~\bibnamefont
  {Liu}}, \bibinfo {author} {\bibfnamefont {M.}~\bibnamefont {Feng}}, \ and\
  \bibinfo {author} {\bibfnamefont {K.}~\bibnamefont {Wang}},\ }\href
  {https://doi.org/10.1103/physreva.84.062109} {\bibfield  {journal} {\bibinfo
  {journal} {Physical Review A}\ }\textbf {\bibinfo {volume} {84}},\ \bibinfo
  {pages} {062109} (\bibinfo {year} {2011})}\BibitemShut {NoStop}%
\bibitem [{\citenamefont {Larson}(2012)}]{Lar:12}%
  \BibitemOpen
  \bibfield  {author} {\bibinfo {author} {\bibfnamefont {J.}~\bibnamefont
  {Larson}},\ }\href {https://doi.org/10.1103/physrevlett.108.033601}
  {\bibfield  {journal} {\bibinfo  {journal} {Physical Review Letters}\
  }\textbf {\bibinfo {volume} {108}},\ \bibinfo {pages} {033601} (\bibinfo
  {year} {2012})}\BibitemShut {NoStop}%
\bibitem [{\citenamefont {Wang}\ \emph {et~al.}(2015)\citenamefont {Wang},
  \citenamefont {Wei},\ and\ \citenamefont {Liang}}]{Wan.Wei.Lia:15}%
  \BibitemOpen
  \bibfield  {author} {\bibinfo {author} {\bibfnamefont {M.}~\bibnamefont
  {Wang}}, \bibinfo {author} {\bibfnamefont {L.}~\bibnamefont {Wei}}, \ and\
  \bibinfo {author} {\bibfnamefont {J.~Q.}\ \bibnamefont {Liang}},\ }\href
  {https://doi.org/10.1016/j.physleta.2015.02.006} {\bibfield  {journal}
  {\bibinfo  {journal} {Physics Letters A}\ }\textbf {\bibinfo {volume}
  {379}},\ \bibinfo {pages} {1087} (\bibinfo {year} {2015})}\BibitemShut
  {NoStop}%
\bibitem [{\citenamefont {Gasparinetti}\ \emph {et~al.}(2016)\citenamefont
  {Gasparinetti}, \citenamefont {Berger}, \citenamefont {Abdumalikov},
  \citenamefont {Pechal}, \citenamefont {Filipp},\ and\ \citenamefont
  {Wallraff}}]{Gas.Ber.Abd.Pec.Fil.Wal:16}%
  \BibitemOpen
  \bibfield  {author} {\bibinfo {author} {\bibfnamefont {S.}~\bibnamefont
  {Gasparinetti}}, \bibinfo {author} {\bibfnamefont {S.}~\bibnamefont
  {Berger}}, \bibinfo {author} {\bibfnamefont {A.~A.}\ \bibnamefont
  {Abdumalikov}}, \bibinfo {author} {\bibfnamefont {M.}~\bibnamefont {Pechal}},
  \bibinfo {author} {\bibfnamefont {S.}~\bibnamefont {Filipp}}, \ and\ \bibinfo
  {author} {\bibfnamefont {A.~J.}\ \bibnamefont {Wallraff}},\ }\href
  {https://doi.org/10.1126/sciadv.1501732} {\bibfield  {journal} {\bibinfo
  {journal} {Science Advances}\ }\textbf {\bibinfo {volume} {2}},\ \bibinfo
  {pages} {e1501732} (\bibinfo {year} {2016})}\BibitemShut {NoStop}%
\bibitem [{\citenamefont {Chryssomalakos}\ \emph {et~al.}(2021)\citenamefont
  {Chryssomalakos}, \citenamefont {Guzm\'an-Gonz\'alez}, \citenamefont
  {Hanotel},\ and\ \citenamefont {Serrano-Ens\'astiga}}]{Chr.Guz.Han.Ser:21}%
  \BibitemOpen
  \bibfield  {author} {\bibinfo {author} {\bibfnamefont {C.}~\bibnamefont
  {Chryssomalakos}}, \bibinfo {author} {\bibfnamefont {E.}~\bibnamefont
  {Guzm\'an-Gonz\'alez}}, \bibinfo {author} {\bibfnamefont {L.}~\bibnamefont
  {Hanotel}}, \ and\ \bibinfo {author} {\bibfnamefont {E.}~\bibnamefont
  {Serrano-Ens\'astiga}},\ }\href {\doibase
  https://doi.org/10.1007/s00220-020-03918-7} {\bibfield  {journal} {\bibinfo
  {journal} {Commun.{} Math.{} Phys.{}}\ }\textbf {\bibinfo {volume} {381}},\
  \bibinfo {pages} {735} (\bibinfo {year} {2021})},\ \bibinfo {note}
  {\texttt{arXiv:1909.02592}}\BibitemShut {NoStop}%
\bibitem [{\citenamefont {Aguilar}\ \emph {et~al.}(2020)\citenamefont
  {Aguilar}, \citenamefont {Chryssomalakos}, \citenamefont
  {Guzm\'an-Gonz\'alez}, \citenamefont {Hanotel},\ and\ \citenamefont
  {Serrano-Ens\'astiga}}]{Agu.Chr.Guz.Han.Ser:20}%
  \BibitemOpen
  \bibfield  {author} {\bibinfo {author} {\bibfnamefont {P.}~\bibnamefont
  {Aguilar}}, \bibinfo {author} {\bibfnamefont {C.}~\bibnamefont
  {Chryssomalakos}}, \bibinfo {author} {\bibfnamefont {E.}~\bibnamefont
  {Guzm\'an-Gonz\'alez}}, \bibinfo {author} {\bibfnamefont {L.}~\bibnamefont
  {Hanotel}}, \ and\ \bibinfo {author} {\bibfnamefont {E.}~\bibnamefont
  {Serrano-Ens\'astiga}},\ }\href@noop {} {\bibfield  {journal} {\bibinfo
  {journal} {J.{} Phys.{} A: Math.{} Theor.{}}\ }\textbf {\bibinfo {volume}
  {53}},\ \bibinfo {pages} {065301} (\bibinfo {year} {2020})},\ \bibinfo {note}
  {arXiv:1903.05022}\BibitemShut {NoStop}%
\end{thebibliography}%
%
\end{document}